\documentclass[%
 reprint,
 amsmath,amssymb,
prx,
floatfix,
superscriptaddress
]{revtex4-2}

\usepackage{mathtools}
\usepackage{graphicx}
\usepackage{dcolumn}
\usepackage{bm}

\usepackage[caption=false]{subfig}

\usepackage{algorithm}
\usepackage[noend]{algpseudocode}
\algrenewcommand\algorithmicdo{}

\makeatletter
\renewcommand{\ALG@name}{Procedure}
\makeatother

\usepackage[linktocpage=true,
  colorlinks=true, 
  pdfborder={0 0 0},
  linkcolor=blue,
  citecolor=red,
  filecolor=yellow,
  urlcolor=blue,
  bookmarks,
  pdfauthor={},
]{hyperref}

\usepackage{orcidlink}

\usepackage{ifthen}
\newcounter{is_qcircuit_used}
\newcounter{are_figs_merged}
\newcommand{\argmax}{\mathop{\rm arg~max}\limits}

\begin{document}

\preprint{APS/123-QED}

\title{
Tensor-decomposition technique for qubit encoding of \\ maximal-fidelity Lorentzian orbitals in real-space quantum chemistry
}

\author{Taichi Kosugi\orcidlink{0000-0003-3379-3361}}
\email{kosugi.taichi@gmail.com}
\affiliation{
Quemix Inc.,
Taiyo Life Nihombashi Building,
2-11-2,
Nihombashi Chuo-ku, 
Tokyo 103-0027,
Japan
}

\affiliation{
Department of Physics,
The University of Tokyo,
Tokyo 113-0033,
Japan
}

\author{Xinchi Huang\orcidlink{0000-0002-7547-074X}}
\affiliation{
Quemix Inc.,
Taiyo Life Nihombashi Building,
2-11-2,
Nihombashi Chuo-ku, 
Tokyo 103-0027,
Japan
}

\affiliation{
Department of Physics,
The University of Tokyo,
Tokyo 113-0033,
Japan
}

\author{Hirofumi Nishi\orcidlink{0000-0001-5155-6605}}
\affiliation{
Quemix Inc.,
Taiyo Life Nihombashi Building,
2-11-2,
Nihombashi Chuo-ku, 
Tokyo 103-0027,
Japan
}

\affiliation{
Department of Physics,
The University of Tokyo,
Tokyo 113-0033,
Japan
}

\author{Yu-ichiro Matsushita\orcidlink{0000-0002-9254-5918}}
\affiliation{
Quemix Inc.,
Taiyo Life Nihombashi Building,
2-11-2,
Nihombashi Chuo-ku, 
Tokyo 103-0027,
Japan
}

\affiliation{Quantum Materials and Applications Research Center,
National Institutes for Quantum Science and Technology (QST),
2-12-1 Ookayama, Meguro-ku, Tokyo 152-8550, Japan
}

\affiliation{
Department of Physics,
The University of Tokyo,
Tokyo 113-0033,
Japan
}

\date{\today}

\begin{abstract}
To simulate the real- and imaginary-time evolution of a many-electron system on a quantum computer based on the first-quantized formalism, we need to encode molecular orbitals (MOs) into qubit states for typical initial-state preparation. 
We propose an efficient scheme for encoding an MO as a many-qubit state from a Gaussian-type solution that can be obtained from a tractable solver on a classical computer.
We employ the discrete Lorentzian functions (LFs) as a fitting basis set,
for which we maximize the fidelity to find the optimal Tucker-form state to represent a target MO.
For $n_{\mathrm{prod}}$
three-dimensional LFs, we provide the explicit circuit construction for the state preparation involving $\mathcal{O} (n_{\mathrm{prod}})$ CNOT gates. 
Furthermore, we introduce a tensor decomposition technique to construct a canonical-form state to approximate the Tucker-form state with controllable accuracy.
Rank-$R$ decomposition reduces the CNOT gate count to $\mathcal{O} (R n_{\mathrm{prod}}^{1/3}).$  
We demonstrate via numerical simulations that the proposed scheme is a powerful tool for encoding MOs of various quantum chemical systems, paving the way for first-quantized calculations using hundreds or more logical qubits.
\end{abstract}

\maketitle 

\section{Introduction}
\label{sec:introduction}

Among the various fields in which quantum computation is expected to outperform the classical computation,
quantum chemistry is a particularly active research field where quantum algorithms are developed and run on current hardware.
In fact, the second-quantized formalism for finding ground states and calculating excitation properties \cite{bib:4470, bib:6119, bib:5005, bib:5163, bib:5569, bib:5570, bib:6796, bib:6806, bib:6807}
is suitable for quantum computation not only for sophisticated methodologies \cite{bib:4517, bib:6801} but also for requiring few resources to encode a many-electron state.
Quantum chemistry in the first-quantized formalism (or equivalently quantum chemistry in real space) \cite{bib:5373, bib:5372, bib:5328, bib:5824, bib:5752, bib:5737, bib:5658, bib:6103, bib:6236, bib:6242, bib:6455, bib:6799, bib:6805}, on the other hand,
requires more qubits than in the second-quantized formalism for performing a meaningful calculation on a quantum computer even for a small molecule.
This fact has been hindering realization of first-quantized calculations despite their favorable features compared to second-quantized calculations.
Looking at the recent growth in the number of high-fidelity qubits on hardware \cite{bib:6326, bib:7108} and realized computation using logical qubits \cite{bib:6948}, however,
we can expect quantum computation using hundreds of logical qubits to be available not far in the future.
This situation highlights the need to develop efficient techniques for all the phases throughout the procedure of a first-quantized calculation, or more broadly a grid-based calculation \cite{bib:6802, bib:6803, bib:6804}.

When performing a first-quantized calculation for simulating the real-time dynamics or finding the ground state \cite{bib:5737, bib:6231, bib:6455} of a many-electron system,
one has to prepare an initial state.
The first-quantized formalism for a quantum computer requires explicit antisymmetrization of an initial state for a many-fermion system \cite{bib:4825, bib:5389, bib:6799},
in contrast to the second-quantized formalism.
It is noted that there also exists a work \cite{bib:6800} applicable to symmetrization for a many-boson system.
The efficient antisymmetrization technique proposed by Berry et al.~\cite{bib:5389} assumes that each of the occupied molecular orbitals (MOs), or equivalently the occupied single-electron states, have been encoded in the corresponding data register.
Decisive efficient schemes for encoding individual MOs, however, have not been developed yet.

\begin{figure*}
\begin{center}
\includegraphics[width=12cm]{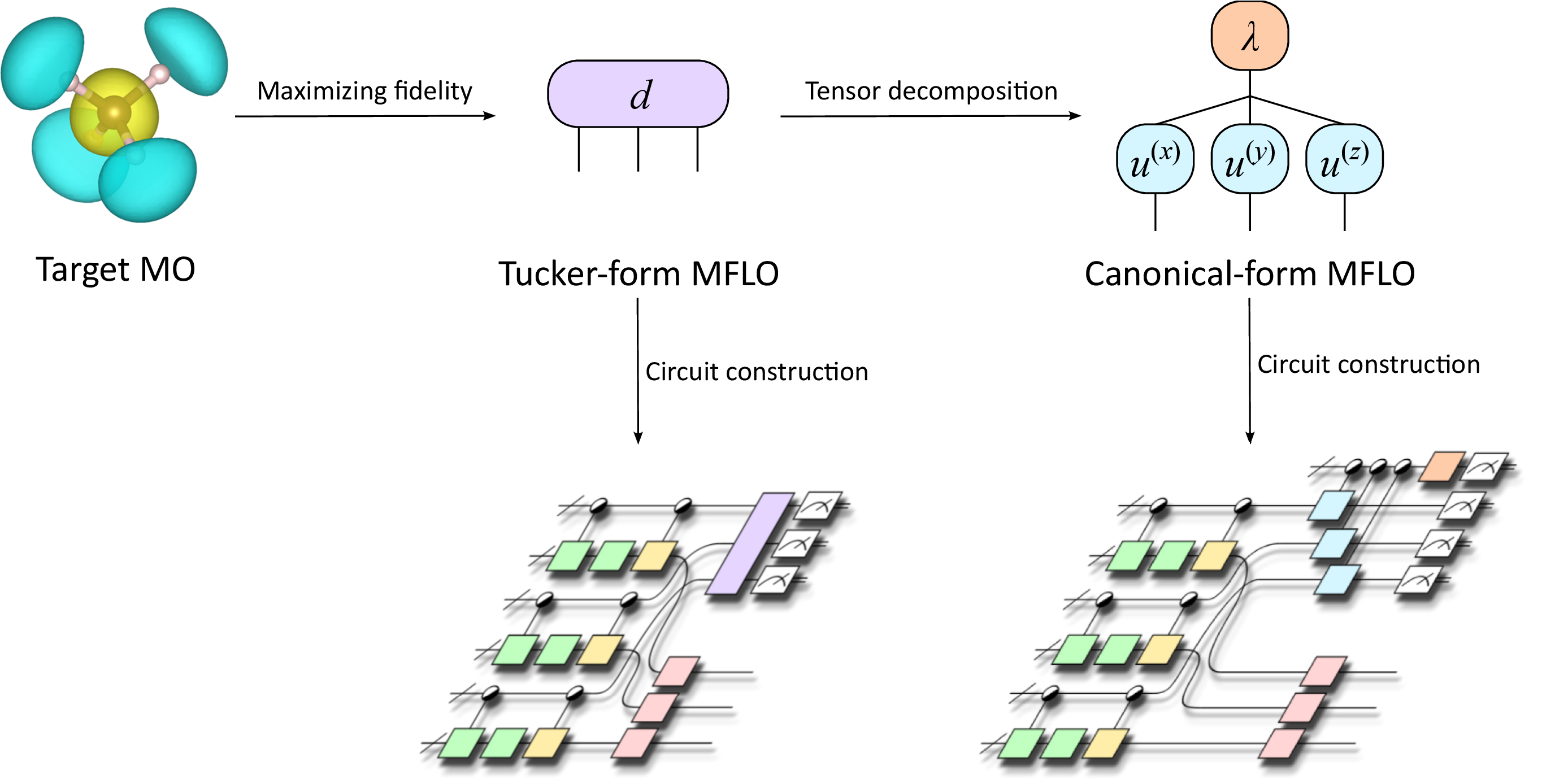}
\end{center}
\caption{
Workflow of the proposed scheme.
For a target MO obtained from a quantum chemistry calculation on a classical computer,
optimal parameters for LFs are found via fidelity maximization to obtain the Tucker-form MFLO.
Tensor decomposition of the core tensor turns the MFLO into a canonical form.
For each of the Tucker- and canonical-forms, efficient circuit implementation for probabilistic qubit encoding is possible.
These circuits can be used for preparing a many-electron state in subsequent quantum computation based on the first-quantized formalism.
}
\label{fig:workflow}
\end{figure*}

The generic scheme for efficient qubit encoding based on the discrete Lorentzian functions (LFs) was recently proposed \cite{bib:6809} based on linear combination of unitaries (LCU).
This scheme is designed for a target state given as a linear combination of localized functions in one- or multi-dimensional space.
We adopt it in the present study and specialize it for practical use of the real-space quantum chemistry by establishing a numerical procedure for constructing maximal-fidelity Lorentzian orbitals (MFLOs),
which can be used as the input to the antisymmetrization technique to prepare a many-electron state.
In addition, we introduce tensor decomposition technique that allows for reduction of gate operations in the circuit to encode an MFLO.
The workflow of our scheme is displayed in Fig.~\ref{fig:workflow}.
We demonstrate via numerical simulations that the proposed scheme is a powerful tool for encoding MOs of various quantum chemical systems. 
It is noted here that there is already a work \cite{bib:6808} for encoding MOs employing tensor train decomposition based on plane-wave basis sets. 
We should keep in mind that
how accurately given MOs are encoded and
how the Slater determinant is suitably composed of those MOs
as a guess for the many-electron ground state
are distinct issues.
Roughly speaking, the former is a topic for quantum information processing, while the latter is that for theories of electronic correlation.

Here we recapitulate briefly initial-state preparation techniques for second-quantized molecular systems as a reference.
Depending on the data format in which an initial second-quantized many-electron state is treated on a classical computer, mainly two types of approaches exist: sum of Slater determinants (SOS) and matrix product state (MPS).
The implementation cost on a quantum computer for an SOS approach is determined mainly by the number of Slater determinants,
while that for an MPS approach by the bond dimension in the tensor train.
SOS and MPS representations can be transformed to each other on a classical computer \cite{bib:7028}.
Various efficient MPS approaches for state preparation have been proposed \cite{bib:7034,bib:7031,bib:5066,bib:7032,bib:7035,bib:7029,bib:7033,bib:7030}.
Meanwhile, there exist systems for which SOS approaches are much more efficient than MPS approaches \cite{bib:7028}.
An SOS approach is efficient when the number of Slater determinants is much fewer than the dimension of the Hilbert space,
while the required resources for an MPS approach are much smaller than SOS approaches thanks to tensor train decomposition.  
For a first-quantized system,
significant amplitudes of a wave function can exist at all points in real space,
indicating that the amplitudes of all the computational bases are basically distributed over the entire Hilbert space.
It thus seems to be difficult to develop first-quantized approaches analogous to SOS approaches.
The generic MPS approaches are, on the other hand,
also applicable to first-quantized systems \cite{bib:6808} as long as classical resources for obtaining the MPS representation of an input wave function are available.

\section{Methods}

\subsection{Preliminaries}

\subsubsection{Gaussian basis functions in continuous space}

Modern quantum chemistry calculation software on classical computers often uses the so-called Cartesian Gaussian (CG) functions \cite{Helgaker}.
The $\mu$th CG function among adopted ones is 
defined to be of the form
\begin{align}
    \chi_{\mu 0} (\boldsymbol{r})
    =
        \overbrace{
            \left(
                \sum_{s}^{n_{\mathrm{G}}}
                b_{\mu s}
                e^{-\gamma_{\mu s} r^2 }
            \right)
        }^{\mathrm{Radial \ part}}
        \overbrace{
            x^{m_{\mu x}}
            y^{m_{\mu y}}
            z^{m_{\mu z}}
        }^{\mathrm{Cartesian \ part}}
        ,
    \label{ampl_encoding_for_MO:def_contracted_Gaussian_AO}
\end{align}
localized at the origin.
The radial part is made up of contracted $n_{\mathrm{G}}$ Gaussian functions having radial exponents
$\{ \gamma_{\mu s} \}_s$ via the coefficients $\{ b_{\mu s} \}_s.$
Each MO in a target molecule is expanded in terms of atomic orbitals (AOs) $\{ \chi_\mu \}_\mu$ that comprise the basis set defined by locating $\{ \chi_{\mu 0} \}_\mu$ at the positions $\{ \boldsymbol{\tau}_\mu \}_\mu$ of individual atoms:
$\chi_\mu (\boldsymbol{r}) \equiv \chi_{\mu 0} (\boldsymbol{r} - \boldsymbol{\tau}_\mu).$
The Cartesian exponents $m_{\mu \nu} \ (\nu = x, y, z)$ in each $\chi_\mu$ specify the orbital angular momentum conveyed by an electron accommodated in the AO.
$\ell_\mu \equiv m_{\mu x} + m_{\mu y} + m_{\mu z}$ is the magnitude of the orbital angular momentum.
While the number of possible combinations of the three exponents is equal to the dimension of multiplet for each of $s$-type ($\ell_\mu = 0$) and $p$-type ($\ell_\mu = 1$) orbitals,
that for $d$-type ($\ell_\mu = 2$) is different from the corresponding dimension of the multiplet.
Specifically, the following six $d$-type CG functions exist: 
$
d_{xx} \propto x^2, \
d_{yy} \propto y^2, \
d_{zz} \propto z^2, \
d_{xy} \propto xy, \
d_{yz} \propto yz, \
d_{zx} \propto zx,
$
despite the five-dimensional irreducible representation of the rotation group. 
That is similarly the case for higher ($\ell_\mu > 2$) orbital angular momenta.
Although this fact makes a basis set constructed from CG functions redundant,
it brings about a striking feature, that is,
each of the $n_{\mathrm{G}}$ terms in 
Eq.~(\ref{ampl_encoding_for_MO:def_contracted_Gaussian_AO})
is factorizable with respect to the three directions as
$
\exp(-\gamma_{\mu s} x^2) x^{m_{\mu x}}
\cdot
\exp(-\gamma_{\mu s} y^2) y^{m_{\mu y}}
\cdot
\exp(-\gamma_{\mu s} z^2) z^{m_{\mu z}}
.
$
It is already known that such factorizability in combination with tensor decomposition techniques allows one to efficiently perform Hartree--Fock calculations on a classical computer 
\cite{bib:6637, bib:6629}.
We will see in the present study that the factorizability is also favorable for efficiently encoding of MOs.

In what follows, we assume that the coefficients of contracted Gaussian functions in $\chi_{\mu 0}$ are normalized in infinite continuous space such that the squared norm is unity: $\int d^3 r |\chi_{\mu 0} (\boldsymbol{r})|^2 = 1.$

\subsubsection{Circuit for a generic LCU}

\begin{figure*}
\begin{center}
\includegraphics[width=16cm]{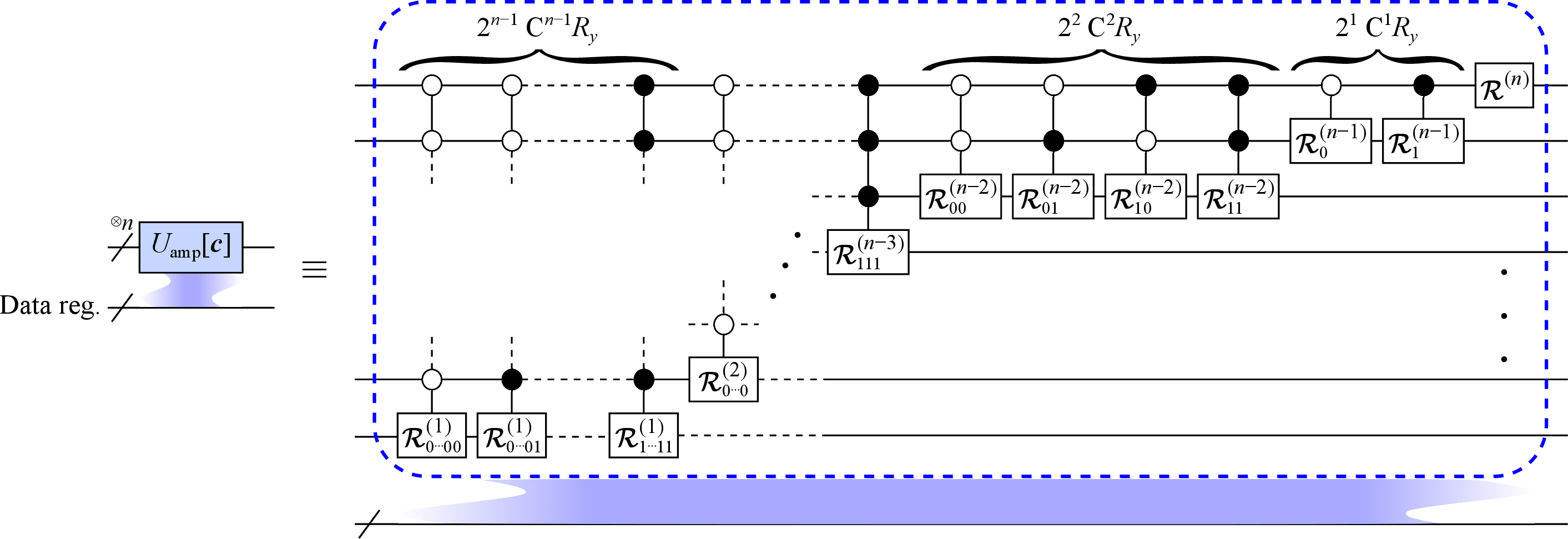}
\end{center}
\caption{
Unitary circuit $U_{\mathrm{amp}}  [\boldsymbol{c}]$ used for generic relative-amplitude encoding of an $n$-qubit system in
Eq.~(\ref{ampl_encoding_of_GMO:ampl_encoding_for_ancillae}).
The single-qubit $y$ rotation gate
$\mathcal{R}^{(m)}_{\lambda} \equiv R_y (-2 \theta^{(m)}_\lambda ),$
where $\lambda$ is empty or a bit string,
is used in this figure.
The explicit expressions for the rotation angles
$\theta^{(m)}_\lambda$ as functions of the amplitudes $\boldsymbol{c}$ are provided in the original paper \cite{bib:5163}.
This circuit is assumed to be used with a data register entangled with the $n$-qubit system from the beginning.
}
\label{fig:circuit_lcu_ampl}
\end{figure*}

The qubit encoding technique \cite{bib:6809} that the present study is based on employs a generic LCU to a many-qubit system \cite{bib:5693, bib:5163}.
The relevant part of the scheme for the present study is the unitary operation $U_{\mathrm{amp}}  [\boldsymbol{c}]$ to generate desired relative amplitudes $\boldsymbol{c}$ on the computational bases of an $n$-qubit system entangled with a data register as
\begin{gather}
    U_{\mathrm{amp}}  [\boldsymbol{c}]
    \sum_{j = 0}^{2^n - 1}
        | \psi_j \rangle 
        | j \rangle_n
    \nonumber \\
    \propto
        \sum_{j = 0}^{2^n - 1}
            c_j
            | \psi_j \rangle 
            | 0 \rangle_n
        +
            (\mathrm{other \ states})
            ,
    \label{ampl_encoding_of_GMO:ampl_encoding_for_ancillae}
\end{gather}
where $| \psi_j \rangle$ is an arbitrary state of the data register.
Since this operation implements amplitude encoding probabilistically in the most generic way,
it requires an exponential cost with respect to $n.$
Composed of many multiply controlled $y$ rotation gates, the circuit is depicted in Fig.~\ref{fig:circuit_lcu_ampl}.
Since the expressions for the circuit parameters are known for given amplitudes $\boldsymbol{c}$,
the success probability can also be calculated on a classical computer.

\subsubsection{Discrete LFs for qubit encoding}

The qubit encoding technique based on discrete LFs \cite{bib:6809} was proposed so that
a given linear combination of localized functions in continuous space
is mapped to grid points represented by many-qubit states.
For an $n$-qubit system,
the normalized discrete LF localized at the origin in one-dimensional (1D) space is defined as
\begin{align}
    L_k (n, a)
    \equiv
        \frac{C_{S} (n, a) }{\sqrt{N}}
        \frac{(1 - e^{-2 a}) (1 - (-1)^k e^{-a N/2})}
        { 1 - 2 e^{-a} \cos (2 \pi k/N) + e^{-2 a} }
        ,
        \label{ampl_encoding_of_GMO:def_discrete_Lorentzian}
\end{align}
for $N \equiv 2^n$ and $k = 0, \dots, N - 1.$
$a$ is the width and $C_{S} (n, a)$ is the normalization constant.
The normalized LF state centered at an integer coordinate $k_{\mathrm{c}}$ is defined as
\begin{align}
    | L ; a, k_{\mathrm{c}} \rangle
    \equiv
        \sum_{k = 0}^{N - 1}
        L_{k - k_{\mathrm{c}}} (n, a)
        | k \rangle_n
    ,
    \label{gen_loc_state:def_Lorentzian_func_state}
\end{align}
where $| k \rangle_n$ is the computational basis.
A linear combination of LFs can be efficiently encoded by first generating the discrete Slater functions (SFs) \cite{bib:5645, bib:6809} with appropriate phase factors
and then performing the quantum Fourier transform (QFT) \cite{Nielsen_and_Chuang}.
For details, see the original paper.

\subsection{MFLO in Tucker form}

\subsubsection{Mapping MO to grid points}

Consider an MO $\phi$ given as a linear combination
\begin{align}
    \phi (\boldsymbol{r})
    =
        \sum_\mu^{n_{\mathrm{bas}}}
        c_\mu
        \chi_{\mu} (\boldsymbol{r})
    \label{ampl_encoding_of_GMO:MO_as_lc_of_AO}
\end{align}
using known MO coefficients $\{ c_\mu \}_\mu$ of the $n_{\mathrm{bas}}$ CG basis functions in continuous space.
Our goal is to amplitude encode $\phi$ as a many-qubit state as accurately as possible under practical restrictions.
To this end, we define a simulation cell as a cube having a sufficiently large edge length $L.$
While we assume the simulation cell to be a cube for simple explanations below,
the edges can have distinct lengths.
We locate the cube at $\boldsymbol{r}_{\mathrm{orig}}$ so that it contain the entirety of the target molecule. 
We then introduce $n_{q e}$ qubits for each direction to define 
$N_{q e} \equiv 2^{n_{q e}}$ equidistant grid points in each direction in the cube.
The spacing between neighboring points is $\Delta x \equiv L/N_{q e}.$
The computational basis
$| \boldsymbol{k} \rangle_{3 n_{q e}}$ specified by three integers $\boldsymbol{k}$
then encodes the position eigenstate of an electron at
$
\boldsymbol{r}^{(\boldsymbol{k})}
\equiv
\boldsymbol{r}_{\mathrm{orig}}
+
\Delta x
(k_x \boldsymbol{e}_x + k_y \boldsymbol{e}_y + k_z \boldsymbol{e}_z)
.
$
From $h (\xi; \gamma, m) \equiv \xi^m e^{-\gamma \xi^2}$ and
$
\widetilde{\boldsymbol{\tau}}_\mu
\equiv
\boldsymbol{\tau}_\mu - \boldsymbol{r}_{\mathrm{orig}}
,
$
we define
\begin{align}
    h_{\mu s}^{(\boldsymbol{k})}
    \equiv
        \prod_{\nu = x, y, z}
        h
        (k_\nu \Delta x - \widetilde{\tau}_{\mu \nu}; \gamma_{\mu s}, m_{\mu \nu})
    ,
\end{align}
where $\gamma_{\mu s}$ and $m_{\mu \nu}$ are the radial exponent and the Cartesian exponent of $\chi_{\mu 0},$ respectively.
[See Eq.~(\ref{ampl_encoding_for_MO:def_contracted_Gaussian_AO})].
The $3 n_{q e}$-qubit state representing the MO inside the simulation cell is then written as
\begin{align}
    | \phi_{\mathrm{ideal}} \rangle
    &\equiv
        \mathcal{N}
        \sqrt{\Delta V}
        \sum_{\boldsymbol{k}}
        \phi
        (\boldsymbol{r}^{(\boldsymbol{k})} + \boldsymbol{r}_{\mathrm{orig}})
        | \boldsymbol{k} \rangle_{3 n_{q e}}
    \nonumber \\
    &=
        \mathcal{N}
        \sqrt{\Delta V}
        \sum_{\boldsymbol{k}}
        \sum_{\mu, s}
        c_\mu
        b_{\mu s}
        h_{\mu s}^{(\boldsymbol{k})}
        | \boldsymbol{k} \rangle_{3 n_{q e}}
    ,
    \label{ampl_encoding_of_GMO:ideal_encoded_phi}
\end{align}
where $\Delta V \equiv (\Delta x)^3$ is the volume element.
\begin{align}
    \mathcal{N}
    \equiv
        \left(
        \Delta V
        \sum_{\boldsymbol{k}}
        |
            \phi
            (\boldsymbol{r}^{(\boldsymbol{k})} + \boldsymbol{r}_{\mathrm{orig}})
        |^2
        \right)^{-1/2}
\end{align}
is a dimensionless constant for
$\langle \phi_{\mathrm{ideal}} | \phi_{\mathrm{ideal}} \rangle = 1$
to hold precisely.
$\mathcal{N}$ is close to 1 when $L$ and $n_{q e}$ are sufficiently large for encoding $\phi.$
The index $\boldsymbol{k}$ of computational basis in
Eq.~(\ref{ampl_encoding_of_GMO:ideal_encoded_phi})
measures the position eigenvalue from $\boldsymbol{r}_{\mathrm{orig}}.$

\subsubsection{Trial state and fidelity}

To establish an efficient scheme for amplitude encoding
the target state $| \phi_{\mathrm{ideal}} \rangle,$
we propose to make use of a Lorentzian basis set.
Let us assume that,
for each direction $\nu = x, y, z,$
the number $n_{\mathrm{L} \nu}$ of available LFs is fixed.
Let $a_{\nu \ell}$ and $k_{\mathrm{c} \nu \ell}$ be the width and center, respectively, of the $\ell$-th LF in the $\nu$ direction.
We then define a trial state as a linear combination of all the possible
$n_{\mathrm{prod}} \equiv n_{\mathrm{L} x} n_{\mathrm{L} y} n_{\mathrm{L} z} $
product basis states,
\begin{gather}
    | \phi_{\mathrm{Tucker}} 
    [\boldsymbol{d}, \boldsymbol{a}, \boldsymbol{k}_{\mathrm{c}}]
    \rangle
    \nonumber \\
    \equiv
        \sum_{\boldsymbol{\ell}}^{n_{\mathrm{prod}}}
            d_{\ell_x \ell_y \ell_z} 
            | L; a_{x \ell_x}, k_{\mathrm{c} x \ell_x} \rangle
            | L; a_{y \ell_y}, k_{\mathrm{c} y \ell_y} \rangle
            | L; a_{z \ell_z}, k_{\mathrm{c} z \ell_z} \rangle
    .
    \label{ampl_encoding_of_GMO:phi_as_Tucker}
\end{gather}
We refer to this state as a Tucker-form trial state since
Eq.~(\ref{ampl_encoding_of_GMO:phi_as_Tucker})
follows the contraction pattern of the Tucker decomposition \cite{bib:6771, bib:4233},
which is one of the most famous tensor decomposition techniques.
The coefficients $\boldsymbol{d}$ in 
Eq.~(\ref{ampl_encoding_of_GMO:phi_as_Tucker})
are called the core tensor.

We need to find the optimal combination of
the core tensor $\boldsymbol{d},$ the widths $\boldsymbol{a},$
and the centers $\boldsymbol{k}_{\mathrm{c}}$ of the LFs.
To this end, we define the fidelity of the Tucker-form trial state as
\begin{align}
    F (\boldsymbol{d}, \boldsymbol{a}, \boldsymbol{k}_{\mathrm{c}})
    \equiv
        |
            \langle \phi_{\mathrm{ideal}} |
            \phi_{\mathrm{Tucker}}
            [\boldsymbol{d}, \boldsymbol{a}, \boldsymbol{k}_{\mathrm{c}}]
            \rangle
        |^2
        -
        P (\boldsymbol{a}, \boldsymbol{k}_{\mathrm{c}})
        ,
    \label{ampl_encoding_of_GMO:def_fidelity}
\end{align}
where the first term on the RHS is the squared overlap between the trial and ideal states.
The second term,
\begin{align}
    P (\boldsymbol{a}, \boldsymbol{k}_{\mathrm{c}})    
    \equiv
        \frac{\alpha_{\mathrm{pen}}}{n_{\mathrm{prod}} }
        \mathrm{Tr}
        \left(
        \left( S (\boldsymbol{a}, \boldsymbol{k}_{\mathrm{c}}) - I
        \right)^2
        \right)
        ,
    \label{ampl_encoding_of_GMO:def_penalty}
\end{align}
introduces a penalty term with a strength parameter $\alpha_{\mathrm{pen}} \geq 0.$
$S (\boldsymbol{a}, \boldsymbol{k}_{\mathrm{c}})$
is the $n_{\mathrm{prod}}$-dimensional overlap matrix between the three-dimensional (3D) LFs appearing in
Eq.~(\ref{ampl_encoding_of_GMO:phi_as_Tucker}) 
and $I$ is the identity matrix.
We maximize the fidelity while obeying the normalization condition
$\| | \phi_{\mathrm{Tucker}} [\boldsymbol{d}, \boldsymbol{a}, \boldsymbol{k}_{\mathrm{c}}] \rangle \|^2 = 1$ to find the MFLO.
The penalty term vanishes if and only if all the 3D LFs are orthogonal to each other.
The larger $\alpha_{\mathrm{pen}}$ is,
the closer the LFs are to an orthonormalized set during the maximization process.
The penalty term may increase the success probability of LCU at the cost of overlap, as will be demonstrated later.

The overlap between the trial and ideal states can be written as
\begin{gather}
    \langle \phi_{\mathrm{ideal}} |
    \phi_{\mathrm{Tucker}}
    [\boldsymbol{d}, \boldsymbol{a}, \boldsymbol{k}_{\mathrm{c}}]
    \rangle
    =
        \sum_{\boldsymbol{\ell}}^{n_{\mathrm{prod}}}
            T_{\boldsymbol{\ell}}
            ( \boldsymbol{a}_{\boldsymbol{\ell}}, \boldsymbol{k}_{\mathrm{c} \boldsymbol{\ell}} )
            d_{\boldsymbol{\ell}}      
    .
    \label{ampl_encoding_of_GMO:ovl_btwn_idel_and_trial}
\end{gather}
The definition of
$
T_{\boldsymbol{\ell}}
( \boldsymbol{a}_{\boldsymbol{\ell}}, \boldsymbol{k}_{\mathrm{c} \boldsymbol{\ell}} )
,
$
which we call the $T$ tensor,
and the derivation of
Eq.~(\ref{ampl_encoding_of_GMO:ovl_btwn_idel_and_trial})
are described in Appendix \ref{sec:derivation_of_ovl}.
Due to the factorizable terms in the AO in 
Eq.~(\ref{ampl_encoding_for_MO:def_contracted_Gaussian_AO})
and the Tucker-form trial state,
the required spatial integrals for evaluating the $T$ tensor are one dimensional.
By using the overlap between two LFs
$
S_{\ell \ell'}^{(\nu)}
(\boldsymbol{a}_\nu, \boldsymbol{k}_{\mathrm{c} \nu} )
\equiv
\langle L; a_{\nu \ell}, k_{\mathrm{c} \nu \ell} |
L; a_{\nu \ell'}, k_{\mathrm{c} \nu \ell'} \rangle
$
in the $\nu$ direction,
we can write the overlap matrix between the product bases as
\begin{align}
    S_{\boldsymbol{\ell}, \boldsymbol{\ell}'} 
    ( \boldsymbol{a}, \boldsymbol{k}_{\mathrm{c}} )
    =
        \prod_{\nu = x, y, z}
        S_{\ell_\nu \ell_\nu'}^{(\nu)}
        (\boldsymbol{a}_\nu, \boldsymbol{k}_{\mathrm{c} \nu} )
        .
    \label{ampl_encoding_of_GMO:def_ovl_mat}
\end{align}
This matrix is clearly symmetric:
$
S_{\boldsymbol{\ell}, \boldsymbol{\ell}'} 
( \boldsymbol{a}, \boldsymbol{k}_{\mathrm{c}} )
=
S_{\boldsymbol{\ell}', \boldsymbol{\ell}} 
( \boldsymbol{a}, \boldsymbol{k}_{\mathrm{c}} )
.
$
The normalization condition to be respected by the trial state is then written as
\begin{gather}
        \sum_{\boldsymbol{\ell}, \boldsymbol{\ell}'}
            d_{\boldsymbol{\ell} }
            S_{\boldsymbol{\ell}, \boldsymbol{\ell}'} 
            ( \boldsymbol{a}, \boldsymbol{k}_{\mathrm{c}} )
            d_{\boldsymbol{\ell}' }
    =
        1
        .
    \label{ampl_encoding_of_GMO:normalization_cond_of_trial}
\end{gather}

\subsubsection{Maximization of fidelity}

For fixed $\boldsymbol{a}$ and $\boldsymbol{k}_{\mathrm{c}},$
the stationarity condition for the fidelity with the optimal core tensor $\widetilde{\boldsymbol{d}}$ becomes 
\begin{align}
    0
    &=
    \left(
        \frac{\partial F (\boldsymbol{d}, \boldsymbol{a}, \boldsymbol{k}_{\mathrm{c}})}{\partial d_{\boldsymbol{\ell} }}
        -
        \kappa
        \frac{
            \partial
            \|
            | \phi_{\mathrm{Tucker}} [\boldsymbol{d}, \boldsymbol{a}, \boldsymbol{k}_{\mathrm{c}}] \rangle
            \|^2
            }{\partial d_{\boldsymbol{\ell}} }
    \right)_{\boldsymbol{d} = \widetilde{\boldsymbol{d}} }
    \nonumber \\
    &=
        \sum_{\boldsymbol{\ell}'}
        \left(
            2
            G_{\boldsymbol{\ell}, \boldsymbol{\ell}'}
            ( \boldsymbol{a}, \boldsymbol{k}_{\mathrm{c}} )
            \widetilde{d}_{\boldsymbol{\ell}'}      
            -
            2
            \kappa
            S_{\boldsymbol{\ell}, \boldsymbol{\ell}'} 
            ( \boldsymbol{a}, \boldsymbol{k}_{\mathrm{c}} )
        \right)
            \widetilde{d}_{\boldsymbol{\ell}' }
        ,
    \label{ampl_encoding_of_GMO:stationarity_cond_for_d}
\end{align}
where we defined the matrix
$G ( \boldsymbol{a}, \boldsymbol{k}_{\mathrm{c}} )$ by
\begin{align}
    G_{\boldsymbol{\ell}, \boldsymbol{\ell}'}
    ( \boldsymbol{a}, \boldsymbol{k}_{\mathrm{c}} )
    \equiv
        T_{\boldsymbol{\ell}}
        ( \boldsymbol{a}_{\boldsymbol{\ell}}, \boldsymbol{k}_{\mathrm{c} \boldsymbol{\ell}} )
        T_{\boldsymbol{\ell}'}
        ( \boldsymbol{a}_{\boldsymbol{\ell}'}, \boldsymbol{k}_{\mathrm{c} \boldsymbol{\ell}'} )
        .
    \label{ampl_encoding_of_GMO:def_G_mat}
\end{align}
$\kappa$ is the Lagrange multiplier.
Since the fidelity is quadratic with respect to the core tensor,
the stationarity condition in
Eq.~(\ref{ampl_encoding_of_GMO:stationarity_cond_for_d})
can be rewritten as the following $n_{\mathrm{prod}}$-dimensional generalized eigenvalue problem for the matrices:
\begin{align}
    G ( \boldsymbol{a}, \boldsymbol{k}_{\mathrm{c}} )
    \widetilde{\boldsymbol{d}}
    =
        \kappa
        S ( \boldsymbol{a}, \boldsymbol{k}_{\mathrm{c}} )
        \widetilde{\boldsymbol{d}}
        .
    \label{ampl_encoding_of_GMO:eig_prob_for_lc_coeffs}
\end{align}
For any solution $\widetilde{\boldsymbol{d}}$ of this problem,
Eqs.~(\ref{ampl_encoding_of_GMO:ovl_btwn_idel_and_trial}),
(\ref{ampl_encoding_of_GMO:normalization_cond_of_trial}),
and
(\ref{ampl_encoding_of_GMO:def_G_mat})
allow us to write the corresponding eigenvalue as
\begin{align}
    \kappa
    &=
    \widetilde{\boldsymbol{d}}
    \cdot
    G ( \boldsymbol{a}, \boldsymbol{k}_{\mathrm{c}} )
    \widetilde{\boldsymbol{d}}
    \nonumber \\
    &=
        \left(
            \sum_{\boldsymbol{\ell}}
            T_{\boldsymbol{\ell}}
            ( \boldsymbol{a}_{\boldsymbol{\ell}},
            \boldsymbol{k}_{\mathrm{c} \boldsymbol{\ell}} )
            \widetilde{d}_{\boldsymbol{\ell}}
        \right)^2
    \nonumber \\
    &=
        F
        (
            \widetilde{\boldsymbol{d}},
            \boldsymbol{a}, \boldsymbol{k}_{\mathrm{c}}
        )
        +
        P (\boldsymbol{a}, \boldsymbol{k}_{\mathrm{c}})
        ,
    \label{ampl_encoding_of_GMO:ovl_btwn_MO_and_trial_using_eigenvector}
\end{align}
which indicates that the eigenvalue is equal to the fidelity plus the penalty.
Let 
$
\{
\kappa_p
( \boldsymbol{a}, \boldsymbol{k}_{\mathrm{c}} ),
\widetilde{\boldsymbol{d}}_p
( \boldsymbol{a}, \boldsymbol{k}_{\mathrm{c}} )
\}_{p = 0}^{n_{\mathrm{prod}} - 1}
$
be the combinations of the eigenvalues and corresponding eigenvectors of this problem.
Although every combination among them satisfies 
Eq.~(\ref{ampl_encoding_of_GMO:ovl_btwn_MO_and_trial_using_eigenvector}),
we would like to find the combination that achieves the highest fidelity.
Since the second term on the right-hand side of
Eq.~(\ref{ampl_encoding_of_GMO:ovl_btwn_MO_and_trial_using_eigenvector})
does not depend on $\widetilde{\boldsymbol{d}},$
the first term achieves its maximum when $\kappa$ is maximized.
Specifically,
the globally optimal core tensor
$\boldsymbol{d} ( \boldsymbol{a}, \boldsymbol{k}_{\mathrm{c}} )$
for the fixed $\boldsymbol{a}$ and $\boldsymbol{k}_{\mathrm{c}}$ is nothing but
the eigenvector belonging to the largest eigenvalue
$
\kappa_{\mathrm{max}}
( \boldsymbol{a}, \boldsymbol{k}_{\mathrm{c}} )
\equiv
\max_p
\kappa_p
( \boldsymbol{a}, \boldsymbol{k}_{\mathrm{c}} )
.
$
That is,
\begin{align}
    \boldsymbol{d} ( \boldsymbol{a}, \boldsymbol{k}_{\mathrm{c}} )
    =
        \widetilde{\boldsymbol{d}}_p
        ( \boldsymbol{a}, \boldsymbol{k}_{\mathrm{c}} )
        \
        \mathrm{with}
        \
        p
        =
        \argmax_p
            \kappa_p
            ( \boldsymbol{a}, \boldsymbol{k}_{\mathrm{c}} )
        .
    \label{ampl_encoding_of_GMO:optimal_d_for_fixed_widths_and_centers}
\end{align}
This core tensor achieves the fidelity
\begin{align}
    F (\boldsymbol{a}, \boldsymbol{k}_{\mathrm{c}} )
    \equiv
    F
    (
    \boldsymbol{d} ( \boldsymbol{a}, \boldsymbol{k}_{\mathrm{c}} ),
    \boldsymbol{a}, \boldsymbol{k}_{\mathrm{c}}
    )
    =
    \kappa_{\mathrm{max}}
    ( \boldsymbol{a}, \boldsymbol{k}_{\mathrm{c}} )
    -
    P (\boldsymbol{a}, \boldsymbol{k}_{\mathrm{c}})
    .
\end{align}
Considering the classical-computational cost $\mathcal{O} (n_{\mathrm{prod}}^{1/3} N_{qe})$ with respect to $n_{\mathrm{prod}}$ and $N_{q e}$ for evaluating the $T$ tensor,
that for obtaining the fidelity in the equation above is
$\mathcal{O} (\max (n_{\mathrm{prod}}^3, n_{\mathrm{prod}}^{1/3} N_{qe})).$
This efficient scaling is because
$n_{\mathrm{prod}}^3$ does not involve the size of data register and 
$n_{\mathrm{prod}}^{1/3} N_{qe}$ has the low scaling thanks to the separability of CG functions and 3D LFs.

Next, we need to maximize the reduced fidelity $F (\boldsymbol{a}, \boldsymbol{k}_{\mathrm{c}} ),$
which no longer involves the variable $\boldsymbol{d}.$
Despite the lack of an explicit expression for the reduced fidelity,
it is possible to calculate numerically its gradient with respect to $\boldsymbol{a}$ as
\begin{gather}
    \frac{
        \partial
        F (\boldsymbol{a}, \boldsymbol{k}_{\mathrm{c}} )
    }{\partial a_{\nu \ell_\nu } }
    =
        2
        f ( \boldsymbol{a}, \boldsymbol{k}_{\mathrm{c}} )
        g_{\nu \ell_\nu } ( \boldsymbol{a}, \boldsymbol{k}_{\mathrm{c}} )
    \nonumber \\
        -
        \kappa_{\mathrm{max}}
        (\boldsymbol{a}, \boldsymbol{k}_{\mathrm{c}} )
        \boldsymbol{d} ( \boldsymbol{a}, \boldsymbol{k}_{\mathrm{c}} )
        \cdot
        \frac{
            \partial S ( \boldsymbol{a}, \boldsymbol{k}_{\mathrm{c}} )
        }{\partial a_{\nu \ell_\nu } }
        \boldsymbol{d} ( \boldsymbol{a}, \boldsymbol{k}_{\mathrm{c}} )
        -
        \frac{\partial P (\boldsymbol{a}, \boldsymbol{k}_{\mathrm{c}})}{\partial a_{\nu \ell} }
        ,
    \label{ampl_encoding_of_GMO:width_deriv_of_ovl_for_optimal_coeffs}
\end{gather}
where we defined
$
f (\boldsymbol{a}, \boldsymbol{k}_{\mathrm{c}} )
\equiv
\sum_{\boldsymbol{\ell} }
T_{\boldsymbol{\ell}}
( \boldsymbol{a}, \boldsymbol{k}_{\mathrm{c}} )
d_{\boldsymbol{\ell}}
( \boldsymbol{a}, \boldsymbol{k}_{\mathrm{c}} )
$
and
\begin{align}
    g_{\nu \ell_\nu} (\boldsymbol{a}, \boldsymbol{k}_{\mathrm{c}} )
    &\equiv
        \sum_{\boldsymbol{\ell}' }
            \frac{
                \partial
                T_{\boldsymbol{\ell}'}
                ( \boldsymbol{a}, \boldsymbol{k}_{\mathrm{c}} )
            }{\partial a_{\nu \ell_\nu } }
            d_{\boldsymbol{\ell}'}
            ( \boldsymbol{a}, \boldsymbol{k}_{\mathrm{c}} )
        .
\end{align}
The derivation of
Eq.~(\ref{ampl_encoding_of_GMO:width_deriv_of_ovl_for_optimal_coeffs})
is described in Appendix \ref{sec:width_deriv_of_ovl_for_optimal_coeffs}.
From the results of numerical diagonalization for 
Eq.~(\ref{ampl_encoding_of_GMO:eig_prob_for_lc_coeffs})
and the known explicit expressions for the overlap matrix in
Eq.~(\ref{ampl_encoding_of_GMO:def_ovl_mat}) and the $T$ tensor,
we can evaluate all the terms on the RHS of
Eq.~(\ref{ampl_encoding_of_GMO:width_deriv_of_ovl_for_optimal_coeffs}).
We can thus use any gradient-based optimization scheme to maximize
$F (\boldsymbol{a}, \boldsymbol{k}_{\mathrm{c}})$ for fixed $\boldsymbol{k}_{\mathrm{c}}$ which further reduces the fidelity:
$
F (\boldsymbol{k}_{\mathrm{c}})
\equiv
\max_{\boldsymbol{a}} F (\boldsymbol{a}, \boldsymbol{k}_{\mathrm{c}}).
$
We should keep in mind that $F (\boldsymbol{k}_{\mathrm{c}})$
depends on initial values of $\boldsymbol{a}$ in general.

To complete the optimization of the trial state,
we need to maximize $F (\boldsymbol{k}_{\mathrm{c}}).$
Since $\boldsymbol{k}_{\mathrm{c}}$ are discrete variables,
we cannot employ gradient-based optimization.

\subsection{Circuit for encoding a Tucker-form state}

\begin{figure*}
\begin{center}
\includegraphics[width=15cm]{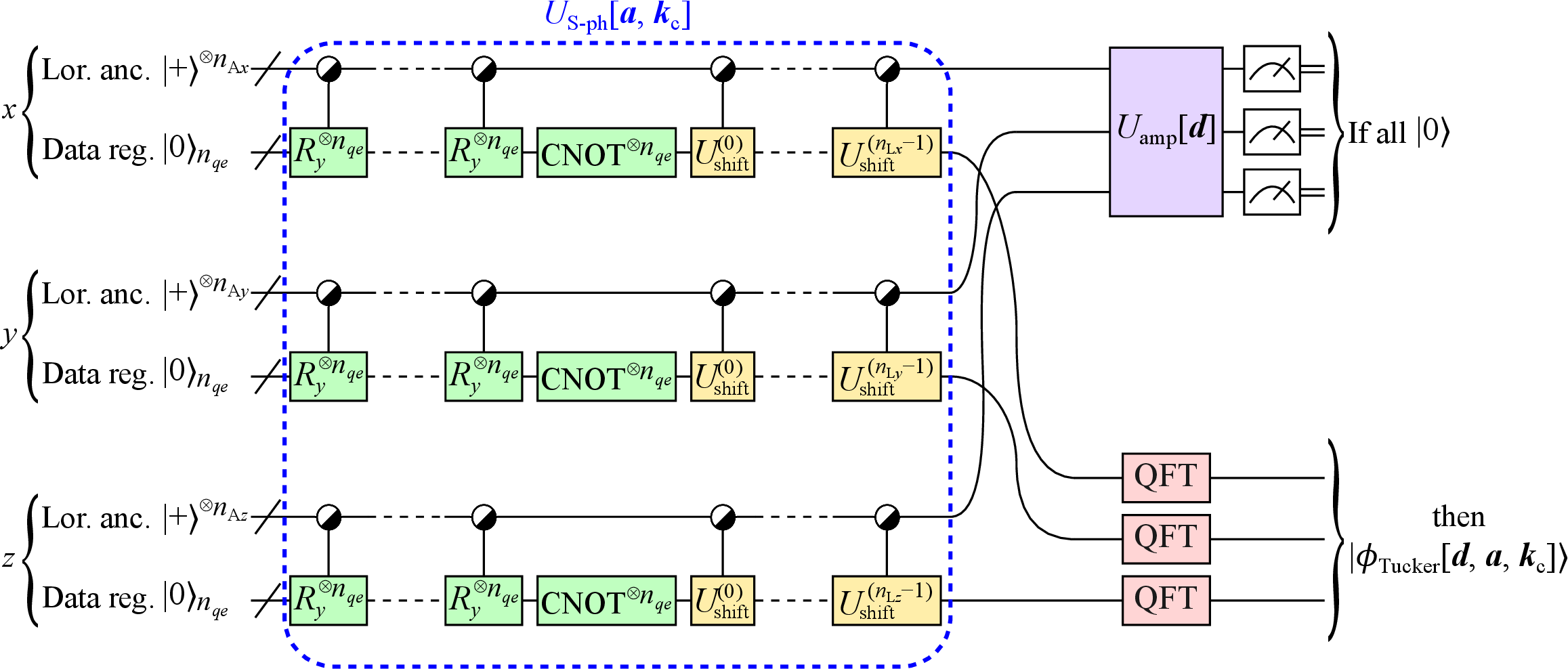}
\end{center}
\caption{
Circuit $\mathcal{C}_{\mathrm{Tucker}}^{(\mathrm{prob})}$
for probabilistic preparation of a Tucker-form state
$
| \phi_{\mathrm{Tucker}} 
[\boldsymbol{d}, \boldsymbol{a}, \boldsymbol{k}_{\mathrm{c}}]
\rangle
.
$
Each of the half-filled circles in this figure represents the multiple control and anticontrol bits.
If the measurement outcome from the
$n_{\mathrm{A}}^{(\mathrm{L})}$
Lorentzian ancillae is all zero,
the desired state has been prepared in the data register.
The dashed region defines the unitary
$U_{\mathrm{S-ph}} [\boldsymbol{a}, \boldsymbol{k}_{\mathrm{c}}]$
for generating the SFs with appropriate phase factors.
}
\label{fig:circuit_tucker}
\end{figure*}

The circuit construction for qubit encoding of a Tucker-form state in 
Eq.~(\ref{ampl_encoding_of_GMO:phi_as_Tucker})
directly uses the techniques proposed in Ref.~\cite{bib:6809}.
To be specific,
the circuit $\mathcal{C}_{\mathrm{Tucker}}^{(\mathrm{prob})}$ for probabilistic preparation is shown in
Fig.~\ref{fig:circuit_tucker}.
We use
\begin{align}
    n_{\mathrm{A} \nu}
    \equiv
        \lceil \log_2 n_{\mathrm{L} \nu} \rceil
    \label{ampl_encoding_of_GMO:num_ancillae_Lorentzian}
\end{align}
Lorentzian ancillae for each direction $\nu$ to discern the ancillary states that designate the indices $\ell_\nu$ of
$| L; a_{\nu \ell_\nu}, k_{\mathrm{c} \nu \ell_\nu} \rangle.$
The total number of ancillae is
$
n_{\mathrm{A}}^{(\mathrm{L})}
\equiv
n_{\mathrm{A} x} + n_{\mathrm{A} y} + n_{\mathrm{A} z} 
$
The partial circuit $U_{\mathrm{S-ph}} [\boldsymbol{a}, \boldsymbol{k}_{\mathrm{c}}]$
generates the SFs with phase factors according to the widths and centers of the LFs,
which are then provided with the desired amplitudes by
$U_{\mathrm{amp}} [\boldsymbol{d}]$
up to overall scaling.
Therefore, the normalized state of the entire system immediately before the measurement is
\begin{align}
        \sqrt{\frac{s_{\mathrm{Tucker}} }{n_{\mathrm{prod}} }}
        | \phi_{\mathrm{Tucker}}
        [\boldsymbol{d}, \boldsymbol{a}, \boldsymbol{k}_{\mathrm{c}}] \rangle
        | 0 \rangle_{ n_{\mathrm{A}}^{(\mathrm{L})} }
        +
        (\mathrm{other \ states})
    ,
\end{align}
where $s_{\mathrm{Tucker}} $ is some value that can be calculated \cite{bib:5163} from the expressions for 
$U_{\mathrm{amp}} [\boldsymbol{d}].$
The success probability for encoding the Tucker-form state is thus
\begin{align}
    \mathbb{P}_{\mathrm{Tucker}}
    =
        \frac{s_{\mathrm{Tucker}} }{n_{\mathrm{prod}}}
        .
    \label{ampl_encoding_of_GMO:success_prob_Tucker}
\end{align}
The QFT operations can be applied to the data register after the measurement is successful since the order of QFT and the measurement does not affect the result of the probabilistic encoding \cite{bib:6809}.

For enhancing the success probability,
quantum amplitude amplification \cite{bib:4884, bib:4878} is applicable.
Since the success probability can be calculated using a classical computer in advance,
we can render it rather close to unity without performing amplitude estimation.
The ratio of the new expected time spent until the completion of the encoding to the old expected time is $\mathcal{O}(1/\sqrt{n_{\mathrm{prod}}})$,
achieving quadratic speedup.
In addition, if an extra ancilla qubit is available, we can make the success probability strictly unity, leading to deterministic encoding \cite{bib:6809}.
This is similarly the case when encoding the canonical-form state introduced below.

\subsection{Canonical-form state from tensor decomposition}

Here we employ the canonical decomposition technique for finding more efficient qubit encoding of a known Tucker-form state.
Since the concept of this generic technique for tensor decomposition was repeatedly reinvented historically,
there are many names for the technique.
For details, see Ref.~\cite{bib:4233} and references therein.

Here, we assume that we already know the optimal parameters for the MFLO in the Tucker form.
For a specified rank $R,$
the numerical machinery of canonical decomposition allows us to approximate the three-leg core tensor $\boldsymbol{d}$ as
\begin{align}
    d_{\ell_x \ell_y \ell_z}
    \approx
        \sum_{r = 0}^{R - 1}
            v^{(x)}_{r \ell_x}
            v^{(y)}_{r \ell_y}
            v^{(z)}_{r \ell_z}
        ,
    \label{ampl_encoding_of_GMO:CANDECOMP_of_Tucker}
\end{align}
where the RHS is represented by the three two-leg tensors $v^{(\nu)} \ (\nu = x, y, z).$
The RHS of the equation just above has a canonical form of rank $R.$
For a generic tensor, canonical decomposition with a higher rank leads to a more accurate approximation.
While exact decomposition is possible for 
$R = n_{\mathrm{prod}}$ in the present case,
it is desirable to find $R$ as small as possible to achieve practical accuracy.
We refer to $v^{(\nu)}$ as the canonical tensors.

Using the normalization constant
\begin{align}
    N_r^{(\nu)}
    \equiv
       \left(
        \sum_{\ell, \ell'}
            v^{(\nu)}_{r \ell}
            v^{(\nu)}_{r \ell'}
            S_{\ell \ell'}^{(\nu)}
            (\boldsymbol{a}_\nu, \boldsymbol{k}_{\mathrm{c} \nu} )
        \right)^{1/2}
    ,
\end{align}
we define the normalized canonical tensor $u^{(\nu)}$ via
$
u_{r \ell}^{(\nu)}
\equiv
v_{r \ell}^{(\nu)} / N_r^{(\nu)}
.
$
We also define the normalized linear combination of LFs as
\begin{align}
    | \phi_r^{(\nu)} \rangle
    \equiv
        \sum_{\ell = 0}^{n_{\mathrm{L} \nu} - 1}
            u^{(\nu)}_{r \ell}
            | L; a_{\nu \ell}, k_{\mathrm{c} \nu \ell} \rangle
\end{align}
and the canonical coefficients
$\lambda_r \equiv N_r^{(x)} N_r^{(y)} N_r^{(z)}.$
Substituting
Eq.~(\ref{ampl_encoding_of_GMO:CANDECOMP_of_Tucker}) into
Eq.~(\ref{ampl_encoding_of_GMO:phi_as_Tucker}), we obtain the following approximation to the Tucker-form state:
\begin{align}
    |
        \phi_{\mathrm{Tucker}}
        [\boldsymbol{d}, \boldsymbol{a}, \boldsymbol{k}_{\mathrm{c}}]
    \rangle
    &\approx
        \sum_{r = 0}^{R - 1}
            \lambda_r
            \underbrace{
            | \phi_r^{(x)} \rangle
            | \phi_r^{(y)} \rangle
            | \phi_r^{(z)} \rangle
            }_{\equiv | \phi_{\mathrm{canon}, r} \rangle }
    \nonumber \\
    &\equiv
        | \phi_{\mathrm{canon}} \rangle
        .
    \label{ampl_encoding_of_GMO:canonical_form_state_normalized}
\end{align}
We refer to $| \phi_{\mathrm{canon}} \rangle$ as the MFLO in the canonical form of rank $R.$
We can assume the canonical coefficients to be in descending order without loss of generality.

It is noted here that the purpose of introducing the tensor decomposition is different from that in MPS-based encoding techniques.
Specifically, an MPS representation for a many-electron state living in a huge Hilbert space is exploited primarily for data compression so that it is tractable on a classical computer and/or storage requirements are reduced.
The low circuit implementation cost for preparing the approximate state is achieved as a result.
We introduced the tensor decomposition, on the other hand, to the encoding technique primarily for reducing the operation number on the quantum circuit, as explained below.
Since the original tensor $\boldsymbol{d}$ is for the single-electron state,
it requires much fewer resources on a classical computer than a many-electron state does.

\subsection{Circuit for encoding a canonical-form state}

\begin{figure*}
\begin{center}
\includegraphics[width=17cm]{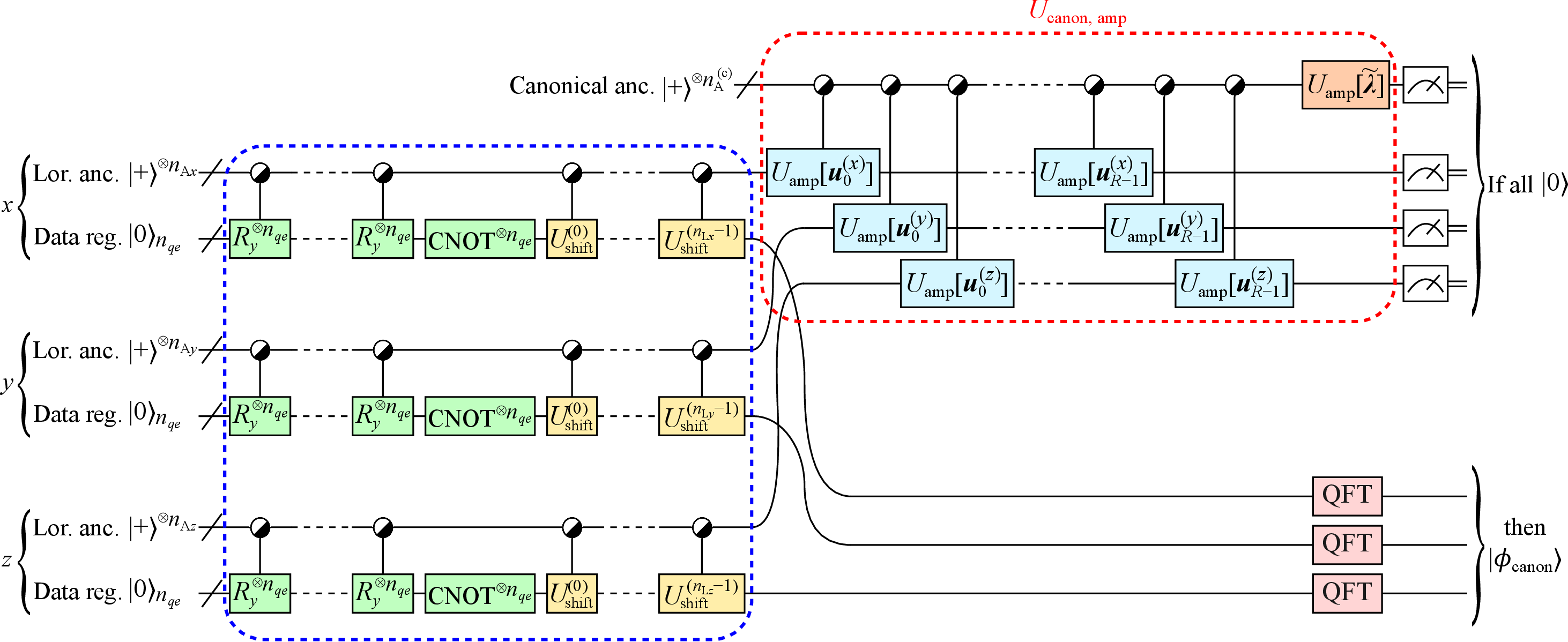}
\end{center}
\caption{
Circuit $\mathcal{C}_{\mathrm{canon}}^{(\mathrm{prob})}$ for probabilistic preparation of a canonical-form state $| \phi_{\mathrm{canon}} \rangle$
originating from a Tucker-form state
$
| \phi_{\mathrm{Tucker}} 
[\boldsymbol{d}, \boldsymbol{a}, \boldsymbol{k}_{\mathrm{c}}]
\rangle
.
$
The blue dashed region indicates the 
$U_{\mathrm{S-ph}} [\boldsymbol{a}, \boldsymbol{k}_{\mathrm{c}}]$ operation,
the same partial circuit as in $\mathcal{C}_{\mathrm{Tucker}}^{(\mathrm{prob})}.$
If the measurement of the Lorentzian and canonical ancillae yields all zeros,
the desired state has been prepared in the data register.
}
\label{fig:circuit_canonical}
\end{figure*}

The circuit for qubit encoding of a canonical-form state in 
Eq.~(\ref{ampl_encoding_of_GMO:canonical_form_state_normalized})
can be constructed by modifying the techniques in
Ref.~\cite{bib:6809}
while considering the contraction pattern for the canonical coefficients and the canonical tensors.
The circuit $\mathcal{C}_{\mathrm{canon}}^{(\mathrm{prob})}$ for probabilistic preparation is shown in
Fig.~\ref{fig:circuit_canonical}.
Besides the Lorentzian ancillae in 
Eq.~(\ref{ampl_encoding_of_GMO:num_ancillae_canonical}),
we use
\begin{align}
    n_{\mathrm{A}}^{(\mathrm{c})}
    \equiv
        \lceil \log_2 R \rceil
    \label{ampl_encoding_of_GMO:num_ancillae_canonical}
\end{align}
canonical ancillae to discern the $R$ terms in the canonical decomposition of the core tensor.

The normalized state of the entire system immediately after the relative-amplitude encoding gates for the normalized canonical tensors $u$ (blue boxes in Fig.~\ref{fig:circuit_canonical}), is
\begin{gather}
        \frac{1}{\sqrt{R n_{\mathrm{prod}} }}
        \sum_r
        \left(
        \bigotimes_{\nu = x, y, z}
            \sqrt{s_r^{(\nu)}}
            | \phi_r^{(\nu)} \rangle
        \right)
        | 0 \rangle_{n_{\mathrm{A}}^{(\mathrm{L})} }
        | r \rangle_{ n_{\mathrm{A}}^{(\mathrm{c})} }
    \nonumber \\
        + (\mathrm{other \ states})
    ,
    \label{ampl_encoding_of_GMO:state_before_ampl_enc_for_canon_coeffs}
\end{gather}
where $s_r^{(\nu)}$ is some value that can be calculated \cite{bib:5163} from the expressions for 
$U_{\mathrm{amp}} [\boldsymbol{u}_{r}^{(\nu)}]$ for each $r.$
To resolve the discrepancy between the coefficients in the equation above and those desired,
we modify the canonical coefficients as
\begin{align}
    \widetilde{\lambda}_r
    \equiv
        \frac{\lambda_r}{\sqrt{s_r^{(x)} s_r^{(y)} s_r^{(z)} } }
    .
\end{align}
The application of 
$U_{\mathrm{amp}} [\widetilde{\boldsymbol{\lambda}}]$
to the state in
Eq.~(\ref{ampl_encoding_of_GMO:state_before_ampl_enc_for_canon_coeffs})
gives
\begin{gather}
        \sqrt{\frac{s_{\mathrm{canon}} }{R n_{\mathrm{prod}} }}
        \sum_r
            \lambda_r
            | \phi_r^{(x)} \rangle
            | \phi_r^{(y)} \rangle
            | \phi_r^{(z)} \rangle
            | 0 \rangle_{ n_{\mathrm{A}}^{(\mathrm{L})}  }
        | 0 \rangle_{ n_{\mathrm{A}}^{(\mathrm{c})} }
    \nonumber \\
        + (\mathrm{other \ states})
    ,
    \label{ampl_encoding_of_GMO:canon_state_before_measurement}
\end{gather}
where $s_{\mathrm{canon}}$ is some value that can be calculated analytically.
The state of the data register coupled to the all-zero ancillary state in
Eq.~(\ref{ampl_encoding_of_GMO:canon_state_before_measurement})
is $| \phi_{\mathrm{canon}} \rangle,$
meaning that $\mathcal{C}_{\mathrm{canon}}^{(\mathrm{prob})}$ implements the desired probabilistic encoding.
The success probability is thus
\begin{align}
    \mathbb{P}_{\mathrm{canon}}
    =
        \frac{s_{\mathrm{canon}}}{R n_{\mathrm{prod}}}
        .
    \label{ampl_encoding_of_GMO:success_prob_canon}
\end{align}

\subsection{Circuit complexities}

The CNOT gate count in a given circuit is often used to quantify its complexity and susceptibility to errors.
While the circuits depicted in
Figs.~\ref{fig:circuit_tucker} and \ref{fig:circuit_canonical}
are direct applications of the circuit in the original paper \cite{bib:6809},
we rearrange the gate operations to allow for the implementation of a uniformly controlled rotation (UCR) \cite{bib:5693} and a diagonal unitary \cite{bib:5352} to minimize the CNOT gate count.
For details, see Appendix \ref{sec:num_cnot_based_on_ucrs}.
The gate count in the state preparation circuit for a Tucker-form state excluding QFT is
\begin{align}
    N_{\mathrm{C} X}
    (\mathcal{C}_{\mathrm{Tucker}}^{(\mathrm{prob})})
    &=
        2^{n_{\mathrm{A}}^{(\mathrm{L})}} 
        -11
        +
        3 n_{q e}
        \sum_{\nu}
            2^{n_{\mathrm{A} \nu} }
    ,
    \label{ampl_encoding_of_GMO:num_cnots_in_Tucker_total}
\end{align}
while that for a canonical-form state is
\begin{align}
    N_{\mathrm{C} X}
    (\mathcal{C}_{\mathrm{canon}}^{(\mathrm{prob})} )
    &=
        -
        2^{ n_{\mathrm{A}}^{(\mathrm{c})} + 1}
        -
        11
        +
        (3 n_{qe} + 2^{ n_{\mathrm{A}}^{(\mathrm{c})} } )
        \sum_\nu
        2^{n_{\mathrm{A} \nu}}
    .
    \label{ampl_encoding_of_GMO:num_cnots_in_canon_total}
\end{align}

For a typical case where the LFs are defined with
$n_{\mathrm{L} x} = n_{\mathrm{L} y} = n_{\mathrm{L} z},$
the scaling of the gate counts derived above in terms of $R$ and $n_{\mathrm{prod}}$ is
$
N_{\mathrm{C} X}
(\mathcal{C}_{\mathrm{Tucker}}^{(\mathrm{prob})})
=
\mathcal{O} (n_{\mathrm{prod}})
$
and
$
N_{\mathrm{C} X}
(\mathcal{C}_{\mathrm{canon}}^{(\mathrm{prob})} )
=
\mathcal{O} (R n_{\mathrm{prod}}^{1/3})
.
$
The canonical-form state thus reduces the gate count compared to the Tucker-form state when low-rank decomposition of the target state is permissible.
Conversely, high-rank decomposition may require more gates than the Tucker-form state preparation.

While we focus on the CNOT gate counts in the present study,
efficient implementation of controlled unitaries with few
gate counts of $T$ or Toffoli operations have been proposed in literature \cite{bib:6827, bib:6828, bib:6829, bib:6830}.
The implementation of the proposed circuits depends heavily on the multiply controlled operations.
If an elementary gate that acts directly on more than two qubits is available,
the actual circuit depths will be reduced largely from those forced to use CNOT gates.
At a physical level, three-qubit (doubly controlled) iToffoli \cite{bib:6603, bib:6602} for superconducting hardware,
$\mathrm{C}^3 Z$ \cite{bib:7056} and $\mathrm{C}^4 Z$ \cite{bib:7057} for trapped ions,
$\mathrm{C}^2 Z$ \cite{bib:6098} for neutral atoms,
and $\mathrm{C}^2 Z$ \cite{bib:7058} for linear optics have been reported to be realized.
Such physical realization of native multiply controlled gates may also help to construct logical circuits since there exist error-correcting codes for which a logical $\mathrm{C}^2 Z$ operation consists only of physical $\mathrm{C}^2 Z$ operations \cite{bib:7083}.

The circuit depths can be estimated similarly to the previous work \cite{bib:6809}.
Those of $\mathcal{C}_{\mathrm{Tucker}}^{(\mathrm{prob})}$ and
$\mathcal{C}_{\mathrm{canon}}^{(\mathrm{prob})},$ including the QFT operations,
scale linearly in terms of $n_{q e}.$
As for $n_{\mathrm{prod}},$ the depth of $\mathcal{C}_{\mathrm{Tucker}}^{(\mathrm{prob})}$
scales as $\mathcal{O} (n_{\mathrm{prod}} \log n_{\mathrm{prod}}),$
while that of $\mathcal{C}_{\mathrm{canon}}^{(\mathrm{prob})}$ scales as $\mathcal{O} (n_{\mathrm{prod}}^{1/3} \log n_{\mathrm{prod}})$ thanks to the canonical decomposition.
$n_{q e}$ scales typically as $\mathcal{O} (\log n_e)$ in terms of the number $n_e$ of electrons contained in a target molecule \cite{bib:5737}.
$n_{\mathrm{prod}}$ is not directly related to $n_e,$ but depends on how precisely the user wants to reproduce the significant amplitudes of the ideal MO,
as will be demonstrated in the examples.

We mention here nontrivial additional cost for implementing fault-tolerant (FT) versions of the proposed circuits briefly.
As seen in Figs.~\ref {fig:circuit_tucker} and \ref {fig:circuit_canonical},
the circuits contain multiply controlled $y$- and $z$-rotations,
each of which can be decomposed into CNOTs and $y$- and $z$-rotations.
Remembering that CNOT and such rotations form a universal gate set \cite{Nielsen_and_Chuang} and that no error-correcting code admits a universal logical gate set composed only of transversally implementable gates \cite{bib:7036},
the logical versions of the proposed circuits inevitably involve non-transversal operations.
We will therefore have to pay attention to suppressing or avoiding error propagation.
In addition, FT implementation of logical non-Clifford operations for the logical rotations is required.
One of the typical techniques for a logical non-Clifford operation is the magic state distillation \cite{bib:6434,bib:6443},
which can be implemented via FT Clifford operations and FT measurements.
Resource estimation taking non-transversal implementation and non-Clifford operations into account for specific error-correcting codes will thus be in order for the FT versions of the proposed circuits.

\subsection{Adoption in an initial many-electron state}

As will be seen in the examples below,
the fidelity between an ideal state and an encoded state may not be close to 1 for the few LFs compared to the system size.
When such an encoded state is used in an initial guess for imaginary-time evolution (ITE) (see, e.g., Ref.~\cite{bib:5737}),
the deviation of the encoded state from the ideal one may not lead to large additional cost thanks to the exponential decay of the weights of undesired states.
For real-time evolution (RTE), on the other hand,
the deviation can lead to distinctly different dynamics from the ideal one.
One possible workaround for that is to perform a small number of steps of variational or nonvariational ITE before starting the RTE so that the undesired states diminish.

\section{Results and discussion}

For all the molecular systems considered below,
we performed electronic-structure calculations by using PySCF \cite{pyscf}.
We used TensorLy \cite{tensorly} for the canonical decomposition of the core tensors.
In all of the cases below, we fixed the centers of LFs at their initial values.

\subsection{H$_2$ molecule}

\begin{figure}
\begin{center}
\includegraphics[width=6cm]{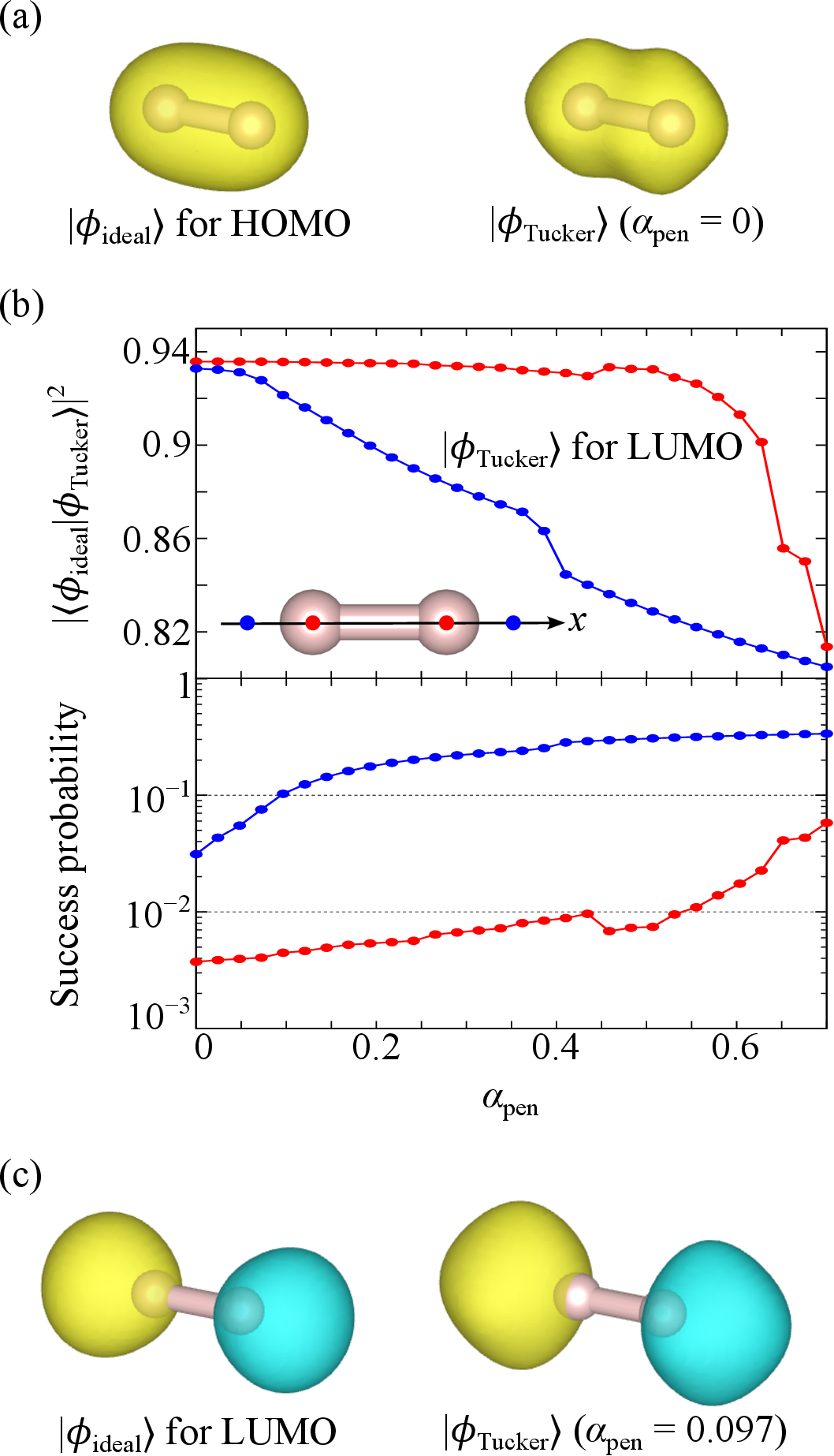}
\end{center}
\caption{
(a)
HOMO of an H$_2$ molecule (left) was used as $| \phi_{\mathrm{ideal}} \rangle$ to be represented as a linear combination of 3D LFs.
The MFLO in the Tucker form (right) encoded as an 18-qubit state was constructed from 3D LFs centered at the two atoms.
This figure was drawn by using VESTA \cite{VESTA}.
(b)
Upper panel shows the deviations of the MFLO from the LUMO as functions of the penalty strength for two patterns of Lorentzian centers.
The red data points are for the centers at the atoms,
while the blue ones are for the centers outside along the bond, as depicted in the inset.
Lower panel shows the success probabilities.
(c)
The LUMO (left) and the corresponding MFLO (right).
}
\label{fig:h2_geom_and_Tucker}
\end{figure}

As the first example, we discuss a hydrogen molecule.
We adopted the 6-31G basis set and the Perdew--Burke--Ernzerhof (PBE) \cite{bib:37} functional for the calculations based on the density functional theory.
We performed qubit encoding of the HOMO and LUMO by using a simulation cell of $L = 8$ a.u. and $n_{q e} = 6$ qubits for the data register in each direction.
We used $n_{\mathrm{L} x} = 2$ LFs for the $x$ direction along the bond,
while $n_{\mathrm{L} y} = n_{\mathrm{L} z} = 1$ for the other directions. 
We adopted two patterns to define the centers of the 1D LFs.
Specifically, the first pattern set the two $x$ centers to the atomic positions,
while the second pattern placed them $0.7$ a.u. away from the atoms outward the bond. [See the inset of Fig.~\ref{fig:h2_geom_and_Tucker}(b)]

The integer coordinates of centers in the first pattern were 
$k_{\mathrm{c} x 0} = 26$ and $k_{\mathrm{c} x 1} = 37$ along the bond,
while those for the other directions were $k_{\mathrm{c} y 0} = k_{\mathrm{c} z 0} = 32.$
The MFLO in the Tucker-form $(\alpha_{\mathrm{pen}} = 0)$ for the HOMO using the first pattern was found to have the dimensionless widths $a_{x 0} = a_{x 1} = 0.643$ and $a_{y 0} = a_{z 0} = 0.744.$
The optimal core tensor $d_{0,0,0} = 0.523$ and $d_{1,0,0} = 0.581$ achieved the squared overlap $0.953$ with the HOMO.
The success probability was $0.82,$ satisfactorily high despite the absence of penalty term in the fidelity.
The discrepancy between the two components of the core tensor despite the reflection symmetry of the diatomic molecule came from the inconsistency between the grid mesh and the atomic positions.
The bonding nature of the HOMO was well reproduced by the MFLO,
as seen in Fig.~\ref{fig:h2_geom_and_Tucker}(a).

The success probabilities for the LUMO were, on the other hand, found to be much lower than for the HOMO, as plotted in Fig.~\ref{fig:h2_geom_and_Tucker}(b).
It was also found that the increase in the penalty strength can remedy the low probabilities at the cost of the squared overlap with the ideal state.
It is also interesting to see that the $x$ centers of LFs away from the bond can give a success probability much higher than for the centers at the atoms.
The MFLO in the Tucker-form $(\alpha_{\mathrm{pen}} = 0.097)$ for the LUMO using the second pattern ($k_{\mathrm{c} x 0} = 20$ and $k_{\mathrm{c} x 1} = 43$) was found to have the dimensionless widths $a_{x 0} = a_{x 1} = 1.34$ and $a_{y 0} = a_{z 0} = 0.960.$
The optimal core tensor $d_{0,0,0} = 1.56$ and $d_{1,0,0} = -1.55$ achieved the squared overlap $0.921$ and the success probability $0.10.$
The nodal structure of the LUMO was well reproduced by the MFLO,
as seen in Fig.~\ref{fig:h2_geom_and_Tucker}(c).
The lower success probability for the LUMO than that for the HOMO might be due to a generic tendency, that is,
an out-of-phase superposition of LFs leads to a lower success probability than an in-phase superposition.
A simple explanation for understanding this tendency is provided in
Appendix \ref{sec:success_prob_lowered_by_antibonding}.

The CNOT gate count (without QFT) for the state preparation circuit $\mathcal{C}_{\mathrm{Tucker}}^{(\mathrm{prob})}$
was 63.
This count came only from the SFs and their phase factors since the amplitude encoding for the core tensor is implemented with no CNOT gate in this case.

\subsection{H$_2$O molecule}

As the second example, we discuss a water molecule.
We adopted the 6-31G basis set and PBE functional.
Among the doubly occupied MOs ($1 a_1, 2 a_1, 1 b_2, 3 a_1,$ and $1 b_1$) of an H$_2$O molecule,
the $1 a_1$ orbital originates almost only from the $1 s$ orbital of the O atom.
We tried qubit encoding of the other four MOs.
We used a simulation cell of $L = 8$ a.u. and $n_{q e} = 7$ for all the calculations below.

\begin{figure*}
\begin{center}
\includegraphics[width=16cm]{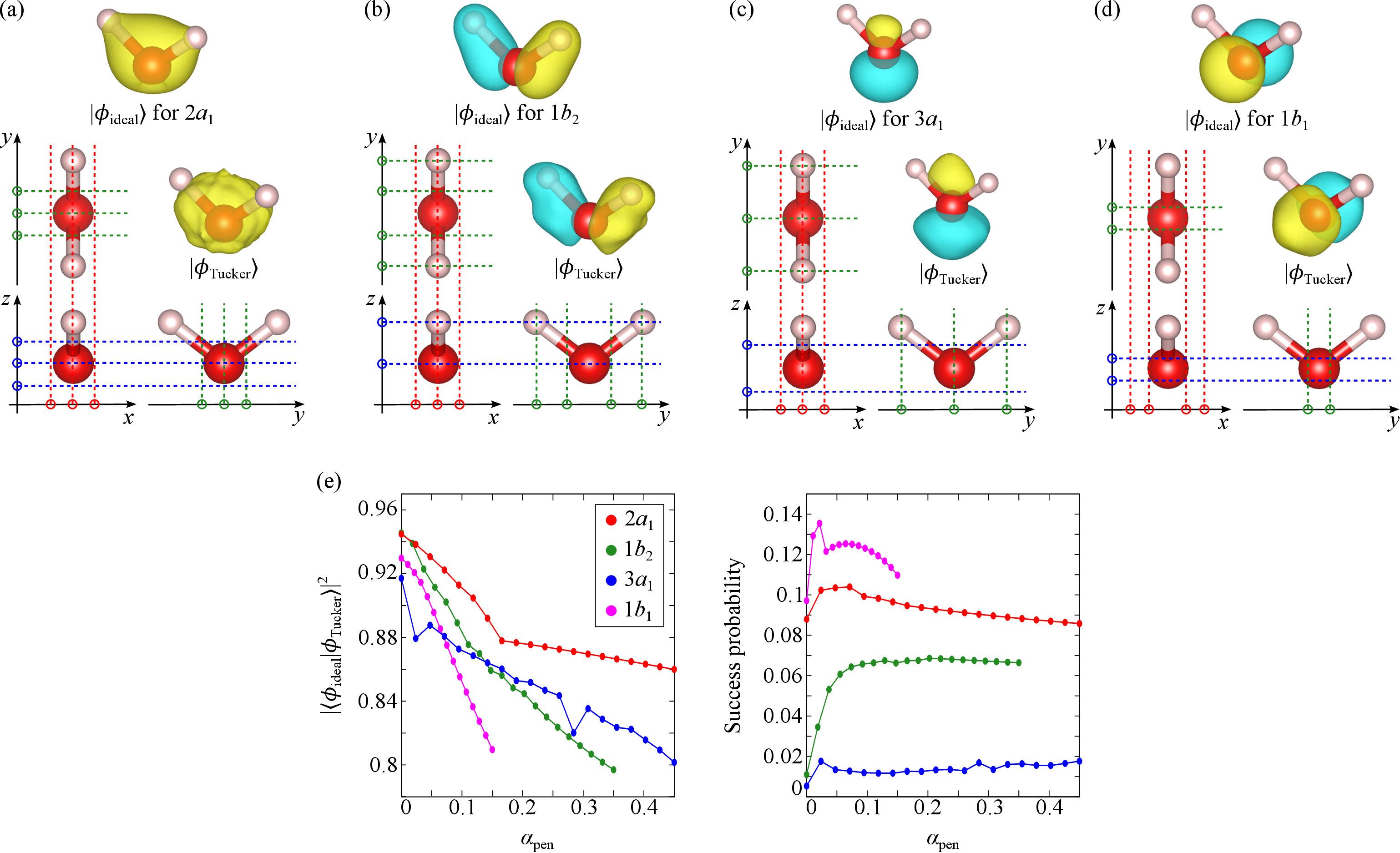}
\end{center}
\caption{
(a) 
$2 a_1$ MO of an H$_2$O molecule is used as $| \phi_{\mathrm{ideal}} \rangle$ to be represented as a linear combination of 3D LFs.
The centers of 1D LFs are designated as circles on the Cartesian axes shown in the orthographic projections.
The MFLO encoded as a 21-qubit state is also shown.
Similar descriptions of $1 b_2, 3 a_1,$ and $1 b_1$ MOs are provided in (b), (c), and (d), respectively.
(e)
Upper panel shows the squared overlaps between the ideal states and the MFLOs as functions of the penalty strength.
Lower panel shows the success probabilities of the state preparation.
}
\label{fig:h2o_geom_and_Tucker}
\end{figure*}

For the individual target MOs,
we defined the centers of 1D LFs as depicted in Figs.~\ref{fig:h2o_geom_and_Tucker}(a)-(d).
We located those centers by seeing the shapes of ideal wave functions.
The MFLOs in the Tucker-form $(\alpha_{\mathrm{pen}} = 0)$ are also shown as 21-qubit states in the figures,
exhibiting the successfully reproduced nodal structures.
Figure \ref{fig:h2o_geom_and_Tucker}(e) shows the squared overlaps between the ideal states and the MFLOs as functions of the penalty strength.
It is found that the introduction of the penalty can increase the success probability for each target MO, as expected.
These probabilities can, however, be lowered as the penalty becomes stronger.
These results tell us that we should find a moderate value of $\alpha_{\mathrm{pen}}$ depending on a target system.

We also performed the tensor decomposition to construct the canonical-form MFLOs from the Tucker-form ones obtained above.
The results are summarized in Table \ref{tab:h2o_cnot_gate_counts},
where the decomposition rank $R$ and the deviation
$1 - |\langle \phi_{\mathrm{Tucker}} | \phi_{\mathrm{canon}} \rangle |^2$
of the canonical-form state from the Tucker-form state for each MO are shown.
We can see that rank-3 decomposition is sufficient for approximating well all the Tucker-form states. 
The estimated CNOT gate counts $N_{\mathrm{C} X}$ are also provided in the table.
It is seen that the benefit of tensor decomposition is small due to the small system size.
For more details of the canonical-form MFLOs, see Appendix \ref{sec:canon_states_of_H2O}.

\begin{table*}[]
\centering
\caption{Numbers $n_{\mathrm{prod}}$ of LFs for product bases fitted to target MOs of an H$_2$O molecule and CNOT gate counts $N_{\mathrm{C} X}$ in the probabilistic state preparation circuits (without QFT) for the Tucker- and canonical-form states. The rank $R$ and the deviation of the canonical-form MFLOs from the Tucker-form ones are also shown. }
\label{tab:h2o_cnot_gate_counts}
\begin{tabular}{cccccc}
\hline
MO      & $n_{\mathrm{prod}} \ (n_{\mathrm{L} x}, n_{\mathrm{L} y}, n_{\mathrm{L} z})$ & $N_{\mathrm{C} X} (\mathcal{C}_{\mathrm{Tucker}}^{(\mathrm{prob})})$ & \multicolumn{1}{l}{$N_{\mathrm{C} X} (U_{\mathrm{S-ph}} [\boldsymbol{a}, \boldsymbol{k}_{\mathrm{c}}])$} & $R$ and deviation        & $N_{\mathrm{C} X} (\mathcal{C}_{\mathrm{canon}}^{(\mathrm{prob})})$ \\ \hline
$2 a_1$ & 27 (3, 3, 3)                                                                 & 305                                                                  & 243                                                                                                      & 3 ($1.1 \times 10^{-6}$) & 281                                                                 \\
$1 b_2$ & 24 (3, 4, 2)                                                                 & 231                                                                  & 201                                                                                                      & 2 ($2.2 \times 10^{-5}$) & 215                                                                 \\
$3 a_1$ & 18 (3, 3, 2)                                                                 & 231                                                                  & 201                                                                                                      & 3 ($6.8 \times 10^{-7}$) & 231                                                                 \\
$1 b_1$ & 16 (4, 2, 2)                                                                 & 173                                                                  & 159                                                                                                      & 2 ($8.5 \times 10^{-8}$) & 169                                                                 \\ \hline
\end{tabular}
\end{table*}

\subsection{Delocalized HOMOs of polyacenes}

\begin{figure*}
\begin{center}
\includegraphics[width=15cm]{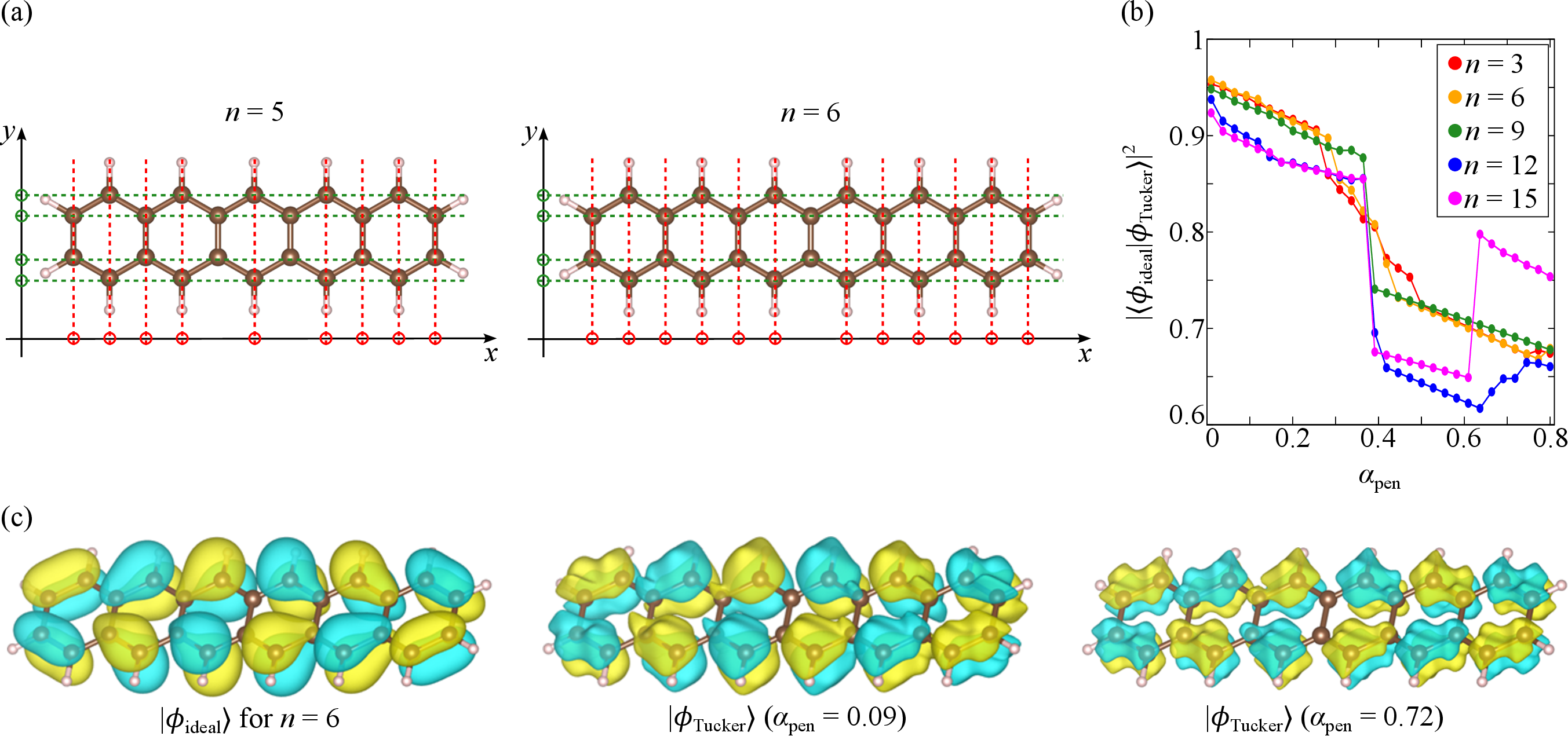}
\end{center}
\caption{
(a)
Centers of LFs for expanding the HOMO of a C$_{4 n + 2}$H$_{2 n + 4}$ molecule composed of $n$ benzene rings.
This figure shows examples for $n = 5$ and 6.
For each molecule, red and green circles represent the centers for $x$ and $y$ directions, respectively,
while the two centers for $z$ direction (not shown) were located at $0.6$ a.u. below and above the molecular plane. 
(b)
Squared overlaps between the ideal states and the optimal Tucker-form states as functions of the penalty strength for various $n.$
(c)
Isosurfaces of HOMO for $n = 6$ as $| \phi_{\mathrm{ideal}} \rangle$ (left),
the MFLO for $\alpha_{\mathrm{pen}} = 0.09$
encoded as a 24-qubit state using the 96 basis functions (middle),
and that for $\alpha_{\mathrm{pen}} = 0.72$ (right).
}
\label{fig:polyacenes_geom_and_Tucker}
\end{figure*}

As the third example, we discuss the HOMOs of polyacenes
C$_{4 n + 2}$H$_{2 n + 4}$ composed of $n$ benzene rings.
We adopted the 6-31G basis set and PBE functional.
For each of $n = 2, \dots, 15,$ we found the HOMO to be delocalized over the entire molecule.
When finding the MFLOs in the Tucker form,
we defined the centers of LFs systematically by referring to the highly symmetric shapes of molecules.
Figure \ref{fig:polyacenes_geom_and_Tucker}(a) shows the centers for
pentacene $(n = 5)$ and hexacene $(n = 6)$ molecules as examples.
For an even (odd) $n,$ our pattern introduced $n_{\mathrm{L} x} = 2 n \ (n_{\mathrm{L} x} = 2 n - 1)$ LFs for the $x$ direction,
while $n_{\mathrm{L} y} = 4$ and $n_{\mathrm{L} z} = 2$ were fixed. 
We used a simulation cell whose edge length $L$ was longer than the molecule length by 7 a.u. for a target $n,$
while $n_{q e} = 8$ was fixed.

Figure \ref{fig:polyacenes_geom_and_Tucker}(b) shows
the squared overlaps between the ideal states and the MFLOs for various ring numbers.
The abrupt changes seen in the overlaps for the varying $\alpha_{\mathrm{pen}}$ may have come from the nonlinear nature of the optimization problem in terms of the widths of LFs.
Figure \ref{fig:polyacenes_geom_and_Tucker}(c) shows the ideal state for $n = 6$ and the corresponding MFLOs.
Although the nodal structure of the ideal wave function is reproduced by the MFLOs for $\alpha_{\mathrm{pen}} = 0.09$ and $0.72,$ the latter exhibits the rugged shape.
This feature has come from the large penalty strengths, which caused effective repulsion between the LFs and forced their widths to be narrow in the optimization process.

\begin{figure*}
\begin{center}
\includegraphics[width=15cm]{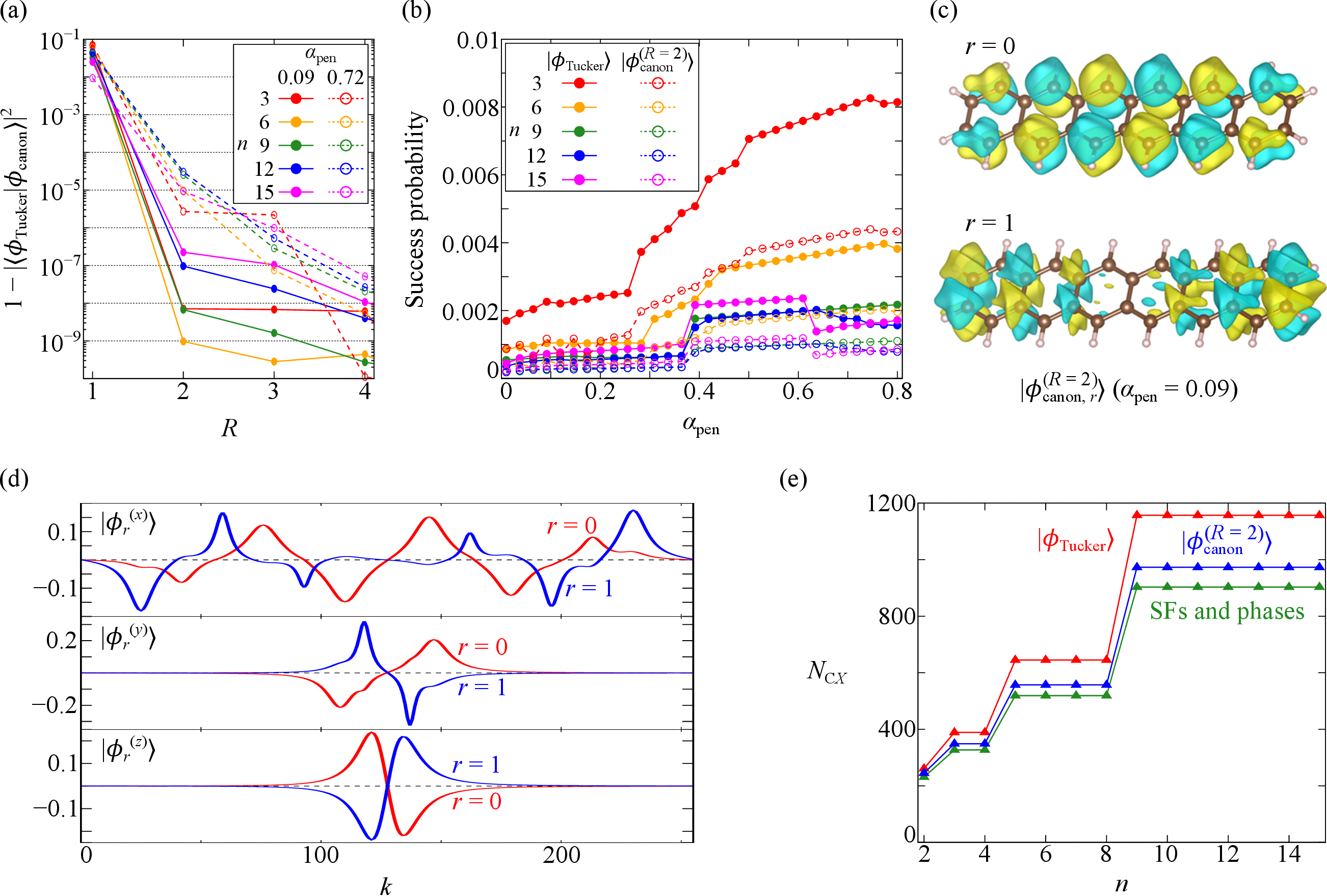}
\end{center}
\caption{
(a)
Deviations of the canonical-form states from the original MFLOs in the Tucker form for C$_{4 n + 2}$H$_{2 n + 4}$ molecules as functions of the rank $R$ for tensor decomposition.
The shown values are for various combinations of the number $n$ of rings and the strength $\alpha_{\mathrm{pen}}$ of penalty.
(b)
Success probabilities for preparing the MFLOs in the Tucker and canonical forms as functions of $\alpha_{\mathrm{pen}}.$
(c)
Two normalized states $| \phi_{\mathrm{canon}, r} \rangle$ comprising the MFLO $| \phi_{\mathrm{canon}} \rangle \ (R = 2)$ in the canonical form for $n = 6.$
(d)
$| \phi_r^{(\nu)} \rangle$ states projected along the computational basis $| k \rangle_{n_{q e}}$ for directions $\nu$.
They constitute $| \phi_{\mathrm{canon}, r} \rangle$ in (c).
(e)
CNOT gate counts for the preparation of the MFLOs as functions of $n.$
Those for the SFs and their phase factors are also shown.
}
\label{fig:polyacenes_canon}
\end{figure*}

Figure \ref{fig:polyacenes_canon}(a) shows the deviations of the canonical-form MFLOs from the original Tucker-form ones as functions of the rank $R$ for tensor decomposition for various combinations of $n$ and $\alpha_{\mathrm{pen}}.$
It is interesting to see that even the canonical-form state for $R = 2$ is in rather good agreement with the ideal state for each case.
The success probabilities for preparing $| \phi_{\mathrm{Tucker}} \rangle$ and
$| \phi_{\mathrm{canon}} \rangle \ (R = 2)$
as functions of $\alpha_{\mathrm{pen}}$ are plotted in Fig.~\ref{fig:polyacenes_canon}(b).
We can see the overall tendency for the probabilities to be higher as the penalty becomes stronger.
While the calculated success probability for $| \phi_{\mathrm{Tucker}} \rangle$ is $7.2 \times 10^{-4},$
that for $| \phi_{\mathrm{canon}} \rangle$ is $3.6 \times 10^{-4}.$
For $n = 6,$ the contributions $| \phi_{\mathrm{canon}, r} \rangle$ to 
the Tucker-form MFLO $(\alpha_{\mathrm{pen}} = 0.09)$ from rank-2 decomposition are depicted in
Fig.~\ref{fig:polyacenes_canon}(c).
We can see that the features of the Tucker-form state are already reproduced well by $| \phi_{\mathrm{canon}, 0} \rangle,$ to which 
$| \phi_{\mathrm{canon}, 1} \rangle$ adds modifications near the two ends of the molecule.
Since each $| \phi_{\mathrm{canon}, r} \rangle$ is separable in terms of the three directions,
it can be plotted as 1D functions, as shown in
Fig.~\ref{fig:polyacenes_canon}(d).
Since there exists only a single node along the $z$ direction in $| \phi_{\mathrm{Tucker}} \rangle,$
the small number of 1D LFs in the $z$ direction suffice to approximate the 
$| \phi_{\mathrm{Tucker}} \rangle.$
It is similarly the case with the $y$ direction.
In contrast, many LFs are needed for the $x$ direction to represent the nodes along the chain of benzene rings.

Fig.~\ref{fig:polyacenes_canon}(e) plots the CNOT gate counts for the preparation of the Tucker-form and rank-2 canonical-form states ss functions of $n.$
The plateau-like behavior of the gate counts is due to the ceiling functions in
Eqs.~(\ref{ampl_encoding_of_GMO:num_ancillae_Lorentzian}) and
(\ref{ampl_encoding_of_GMO:num_ancillae_canonical})
for evaluating
Eqs.~(\ref{ampl_encoding_of_GMO:num_cnots_in_Tucker_total}) and
(\ref{ampl_encoding_of_GMO:num_cnots_in_canon_total}).
The gate counts are found to consist mainly of the process for generating the SFs and their phase factors (see the green points in the figure).
Recalling that this process is common to the Tucker- and canonical-form states,
we can understand that the gate counts coming from the amplitude encoding are reduced drastically by the tensor decomposition.
This benefit comes mainly from the fact that the Tucker-form states allow for the low-rank canonical decomposition in the present system.

\subsection{Localized MOs of iron porphyrin complex}

As the fourth example, we discuss a phenolate-bound iron porphyrin complex (POR).
The ground state of a POR in a non-coordinating solvent is known to contain
the five-coordinate high-spin Fe$^{\mathrm{III}}$ atom \cite{bib:6784, bib:6780} with the total spin $S = 5/2.$
For the element Fe, we used LANL2DZ effective core potential \cite{bib:4439,bib:4440,bib:4441} and its associated basis set.
For the other elements, we adopted the 6-31G* basis set.
We used BP86 \cite{bib:6794,bib:6795} functional.
We set the difference in electron number between the majority- and minority-spin states to 5 for solving the unrestricted Kohn--Sham equation,
that is, we allowed the spatial orbitals of distinct spins to differ from each other.
We used the optimized molecular geometry provided by Das and Dey \cite{bib:6780}.

We tried qubit encoding of the first and fourth highest singly occupied molecular orbitals (SOMO) of the calculated ground state.
The lower-energy MO among the two was found to be derived from
the $a_{1 u}$ MO of a porphine molecule \cite{bib:6797}
and has no significant amplitude around the Fe ion and the phenolate ligand.
We denote this MO as the $a_{1 u}$ SOMO in what follows. 
In contrast, the higher-energy MO was found to have not only the amplitudes around the porphyrin macrocycle,
but also those from the Fe $d_{z^2}$ orbital and the ligand.
We denote this MO as the $d_{z^2}$ SOMO in what follows.
It is noted that the notation $z$ in $d_{z^2}$ is for the direction perpendicular to the molecular plane of the porphyrin macrocycle.
We used a simulation cell of $L = 36$ a.u. and $n_{q e} = 8$ for all the calculations below.

\begin{figure*}
\begin{center}
\includegraphics[width=14cm]{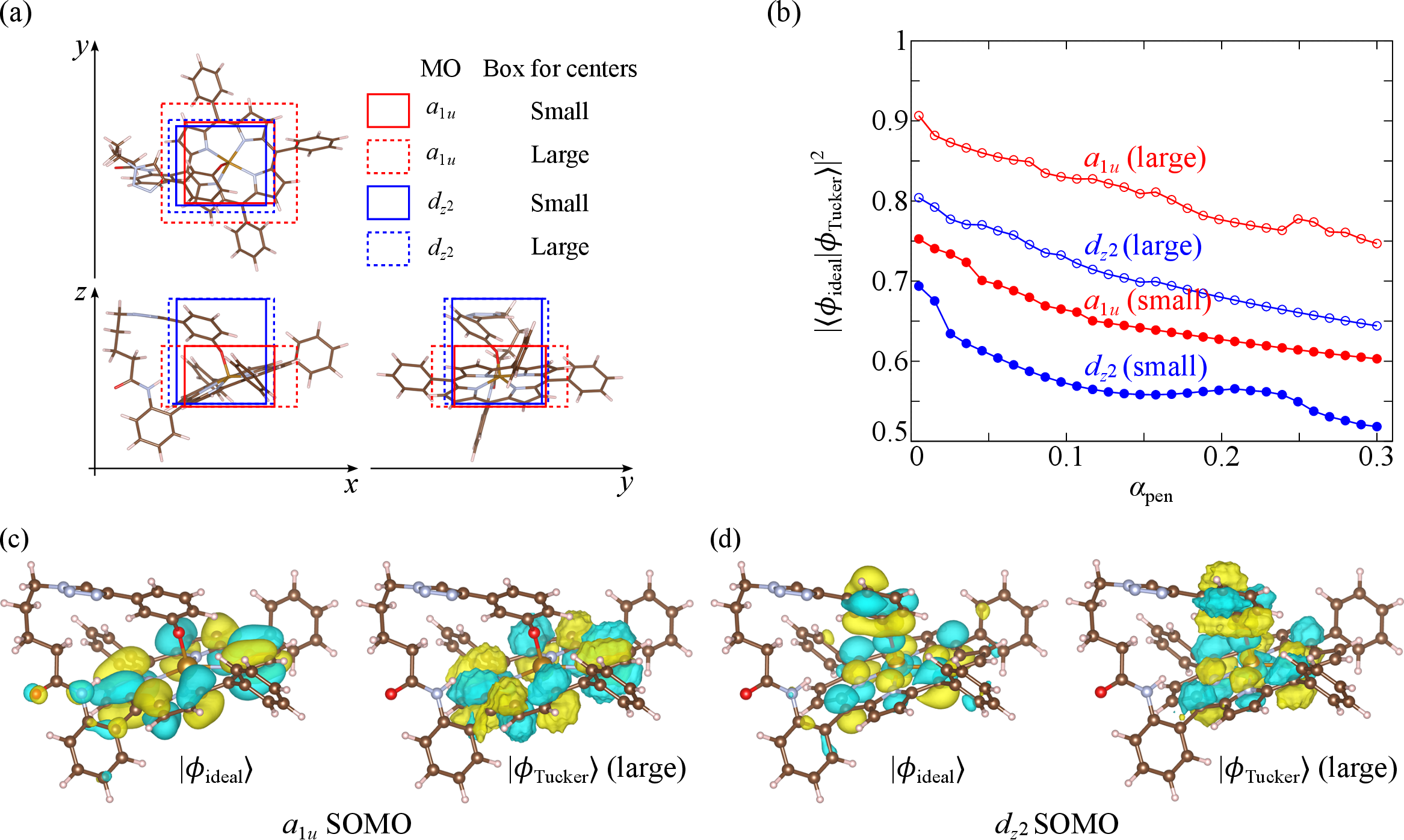}
\end{center}
\caption{
(a)
Orthographic projections of the POR contained in the simulation cell of $L = 36$ a.u.,
where the boxes are for defining the centers of LFs for qubit encoding of $a_{1 u}$ and $d_{z^2}$ SOMOs.
(b)
Squared overlaps between the ideal states and the MFLOs in the Tucker form as functions of the penalty strength when the MOs are encoded using the small and large boxes.
(c)
Isosurfaces of $a_{1 u}$ orbital as $| \phi_{\mathrm{ideal}} \rangle$ (left)
and the MFLO for $\alpha_{\mathrm{pen}} = 0.036$ encoded as a 24-qubit state using the large box (right).
(d)
shows the same for the $d_{z^2}$ SOMO.
}
\label{fig:porphyrin_geom_and_Tucker}
\end{figure*}

The qubit encoding for this system is challenging since the atoms in the POR are not located in highly symmetric positions in contrast to the other example systems discussed above.
We therefore decided to introduce a box for defining the centers of 3D LFs for encoding an MO.
Specifically,
we defined the box inside which the centers are located regularly according to the edge lengths $L_{\mathrm{box}, \nu} \ (\nu = x, y, z)$ of the box.
When encoding the $a_{1 u}$ SOMO,
we used a small box
($L_{\mathrm{box}, x} = L_{\mathrm{box}, y} = 12$ a.u. and $L_{\mathrm{box}, z} = 8$) for 384 basis functions
($n_{\mathrm{L} x} = n_{\mathrm{L} y} = 8$ and $n_{\mathrm{L} z} = 6$)
and a large box
($L_{\mathrm{box}, x} = L_{\mathrm{box}, y} = 18$ a.u. and $L_{\mathrm{box}, z} = 8$)
for 1960 basis functions
($n_{\mathrm{L} x} = n_{\mathrm{L} y} = 14$ and $n_{\mathrm{L} z} = 10$).
When encoding the $d_{z^2}$ SOMO,
we used a small box
($L_{\mathrm{box}, x} = L_{\mathrm{box}, y} = 12$ a.u. and $L_{\mathrm{box}, z} = 14$) 
for 1000 basis functions
($n_{\mathrm{L} x} = n_{\mathrm{L} y} = n_{\mathrm{L} z} = 10$)
and a large box
($L_{\mathrm{box}, x} = L_{\mathrm{box}, y} = L_{\mathrm{box}, z} = 14$)
for 2744 basis functions
($n_{\mathrm{L} x} = n_{\mathrm{L} y} = n_{\mathrm{L} z} = 14$).
Fig.~\ref{fig:porphyrin_geom_and_Tucker}(a)
shows the orthographic projections of the POR and the boxes.

The squared overlaps between the ideal states and the Tucker-form MFLOs for the small and large boxes are plotted in
Fig.~\ref{fig:porphyrin_geom_and_Tucker}(b).
Going through the results we obtained for various settings, including those not shown here,
we found that increase in the basis functions tends to cause larger effects on the overlaps than the extension of the boxes does.
Figure \ref{fig:porphyrin_geom_and_Tucker}(c) and (d)
show the ideal states and the Tucker-form MFLOs $(\alpha_{\mathrm{pen}} = 0.036)$ using the large box for the $a_{1 u}$ and $d_{z^2}$ SOMOs, respectively.
We see that these Tucker-form MFLOs capture the overall features of the ideal states well.
The rugged shapes of the Tucker-form states looking like kompeitos have come from the effective repulsion between the regularly distributed many 3D LFs as well as for polyacenes mentioned above.
The overlap for the $d_{z^2}$ orbital tends to be lower than that for the $a_{1 u}$ orbital [see Fig.~\ref{fig:porphyrin_geom_and_Tucker}(b)].
It may be due to the longer box edge length in the $z$ direction 
and the larger number of nodes of wave function for the $d_{z^2}$ orbital than for the $a_{1 u}$ orbital.

\begin{figure*}
\begin{center}
\includegraphics[width=16cm]{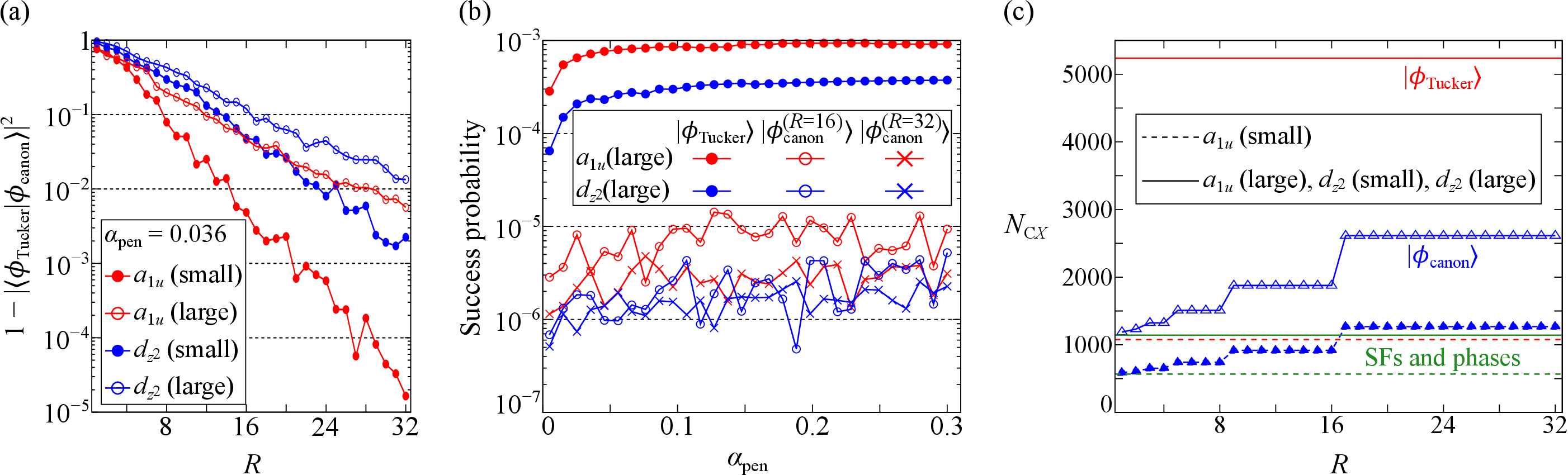}
\end{center}
\caption{
(a)
Deviations of the canonical-form states from the original MFLOs in the Tucker form for the $a_{1 u}$ and $d_{z^2}$ SOMOs of the POR as functions of the rank $R$ for tensor decomposition.
(b)
Success probabilities for preparing the MFLOs in the Tucker and canonical forms as functions of penalty strength.
(c)
CNOT gate counts for the preparation of the MFLOs as functions of $R.$
Those for the SFs and their phase factors are also shown.
}
\label{fig:porphyrin_canon}
\end{figure*}

Figure \ref{fig:porphyrin_canon}(a) shows the deviations of the canonical-form MFLOs from the original Tucker-form ones as functions of the rank $R$ for tensor decomposition for $\alpha_{\mathrm{pen}} = 0.036,$
where it is seen that the large boxes require higher $R$ than the small boxes for achieving high accuracy in the decomposition.
There may exist two main reasons for the necessity of high-rank decomposition.
The first one is simply that the sizes of core tensors in the large-box cases are larger than the small-box cases, that is, the large-box cases involve more basis functions.
The second one is that the target MOs in the POR have shapes that essentially require high-rank decomposition.
The second reason comes from the low-symmetric molecular geometry of the POR,
in contrast to the cases of polyacenes discussed above.
The success probabilities for preparing $| \phi_{\mathrm{Tucker}} \rangle$ and
$| \phi_{\mathrm{canon}} \rangle \ (R = 16 \ \mathrm{and} \ 32)$
as functions of the penalty strength are plotted in Fig.~\ref{fig:porphyrin_canon}(b).
Those for $| \phi_{\mathrm{canon}} \rangle$ are rather lower than for $| \phi_{\mathrm{Tucker}} \rangle.$
The oscillatory behavior of the probabilities is a consequence of the complicated interferences between the many 3D LFs.

Figure \ref{fig:porphyrin_canon}(c) plots the CNOT gate counts for the preparation of the Tucker-form and canonical-form MFLOs as functions of the decomposition rank.
The gate counts for the SFs and phases are constant in the plot since they do not depend on the decomposition rank.
For the $a_{1 u}$ SOMO using the small box, the gate count for preparing
$|\phi_{\mathrm{canon}} \rangle$ exceeds that for $| \phi_{\mathrm{Tucker}} \rangle$
when $R > 16,$ meaning that the high-rank decomposition (compared to $n_{\mathrm{prod}}$) brings about no benefit.
In contrast, for the other cases, the gate count for preparing
$| \phi_{\mathrm{canon}} \rangle$ is approximately half of that for
$| \phi_{\mathrm{Tucker}} \rangle$ even when $R = 32.$
Recalling that the success probabilities for the $| \phi_{\mathrm{canon}} \rangle$ states are roughly $10^{-3}$ to $10^{-2}$ of those for the $| \phi_{\mathrm{Tucker}} \rangle$ states [see Fig.~\ref{fig:porphyrin_canon}(b)],
the expected computational time for completing the state preparation using the canonical-form states is found to be much longer than using the Tucker-form states.
We are thus left with two choices:
the larger gate counts with higher success probabilities (Tucker-form states)
or
the smaller gate counts with lower success probabilities (canonical-form states).
We may be confronted with such choices in a generic case where a target MO requires high-rank tensor decomposition.
Which one should be adopted practically needs to be determined by referring to the coherence time of hardware being used,
which dominates the permissible time for a single run of a circuit.

\section{Conclusions}

In summary,
we proposed an efficient scheme for encoding an MO as a many-qubit state referred to as the MFLO,
numerically constructed via maximization of the fidelity of a trial state.
Since the proposed scheme exploits the fact that a typical MO consists of a small number of localized orbitals, independently of the grid resolution for encoding,
the CNOT gate count is not affected largely by the size of a data register.
We demonstrated that the scheme works and the tensor decomposition technique can reduce the gate count.
This scheme will be a powerful tool for performing quantum chemistry in real space on a quantum computer using hundreds of or more logical qubits in the future.

We mention here the Fourier interpolation scheme \cite{bib:5929, bib:6136} for encoding an MO.
One of its fascinating features against the present scheme is the unitarity \cite{bib:5693, bib:6136}.
For a case in which the determination of circuit parameters 
from the Fourier components on the coarse grid for interpolation can be completed within practical time,
the interpolation scheme will be a good option.

\begin{acknowledgments}
We thank Yannick Couzini\'e for fruitful discussion. This work was supported by Japan Society for the Promotion of Science (JSPS) KAKENHI under
Grant-in-Aid for Scientific Research No.21H04553, No.20H00340, and No.22H01517. This work was
partially supported by the Center of Innovations for Sustainable Quantum AI (JST Grant number
JPMJPF2221). The authors thank the Supercomputer Center, the Institute for Solid State Physics, the Science Tokyo for the use of the facilities.
\end{acknowledgments}

\begin{widetext}

\appendix

\section{Derivation of Eq.~(\ref{ampl_encoding_of_GMO:ovl_btwn_idel_and_trial})}
\label{sec:derivation_of_ovl}

For $\nu = x, y, z,$
we define quantities
\begin{align}
    M_{\nu \mu s} (a, k_{\mathrm{c}})
    &\equiv
        \frac{L}{\sqrt{N_{q e}}}
        \sum_{k = 0}^{N_{q e} - 1}
        h (k \Delta x - \widetilde{\tau}_{\mu \nu}; \gamma_{\mu s}, m_{\mu \nu})
        L_{k - k_{\mathrm{c}} } (n_{q e}, a)
\end{align}
that depend not on the target MO $\phi$, but on the AO $\chi_\mu.$
These are nothing but the numerical integrals of the products of the CG functions and the LFs in 1D space.
$M_{\nu \mu s} (a, k_{\mathrm{c}})$ has the dimension of length$^{m_{\mu \nu} + 1}.$
From them, we define dimensionless quantities 
\begin{align}
    R_{\mu \boldsymbol{\ell}}
    ( \boldsymbol{a}_{\boldsymbol{\ell} }, \boldsymbol{k}_{\mathrm{c} \boldsymbol{\ell}} )
    &\equiv
        \frac{1}{\sqrt{L^3}}
        \sum_s^{n_{\mathrm{G}}}
        b_{\mu s}
            M_{x \mu s} (a_{x \ell_x}, k_{\mathrm{c} x \ell_x})
            M_{y \mu s} (a_{y \ell_y}, k_{\mathrm{c} y \ell_y})
            M_{z \mu s} (a_{z \ell_z}, k_{\mathrm{c} z \ell_z})
        .
\end{align}

From the expression of the ideal state in
Eq.~(\ref{ampl_encoding_of_GMO:ideal_encoded_phi})
and that of the Tucker-form trial state in
Eq.~(\ref{ampl_encoding_of_GMO:phi_as_Tucker}),
we have
\begin{gather}
    \langle \phi_{\mathrm{ideal}} |
    \phi_{\mathrm{Tucker}}
    [\boldsymbol{d}, \boldsymbol{a}, \boldsymbol{k}_{\mathrm{c}}]
    \rangle
    \nonumber \\
    =
        \mathcal{N}
        \sqrt{\Delta V}
        \sum_s^{n_{\mathrm{G}}}
        \sum_\mu^{n_{\mathrm{bas}}}
        c_\mu
        b_{\mu s}
        \sum_{\boldsymbol{k}}
        h (x^{(k_x)} - \widetilde{\tau}_{\mu x}; \gamma_{\mu s}, m_{\mu x})
        h (y^{(k_y)} - \widetilde{\tau}_{\mu y}; \gamma_{\mu s}, m_{\mu y})
        h (z^{(k_z)} - \widetilde{\tau}_{\mu z}; \gamma_{\mu s}, m_{\mu z})
    \cdot
    \nonumber \\
    \cdot
        \sum_{\boldsymbol{\ell}}^{n_{\mathrm{prod}}}
            d_{\ell_x \ell_y \ell_z}      
        L_{k_x - k_{\mathrm{c} x \ell_x} } (n_{q e}, a_{x \ell_x})
        L_{k_y - k_{\mathrm{c} y \ell_y} } (n_{q e}, a_{y \ell_y})
        L_{k_z - k_{\mathrm{c} z \ell_z} } (n_{q e}, a_{z \ell_z})
    \nonumber \\
    =
        \sum_{\boldsymbol{\ell}}^{n_{\mathrm{prod}}}
            T_{\boldsymbol{\ell}}
            ( \boldsymbol{a}_{\boldsymbol{\ell}}, \boldsymbol{k}_{\mathrm{c} \boldsymbol{\ell}} )
            d_{\boldsymbol{\ell}}      
    ,
\end{gather}
where we defined the dimensionless quantity
\begin{align}
    T_{\boldsymbol{\ell}}
    ( \boldsymbol{a}_{\boldsymbol{\ell} }, \boldsymbol{k}_{\mathrm{c} \boldsymbol{\ell}} )
    &\equiv
        \mathcal{N}
        \sum_{\mu}^{n_{\mathrm{bas}}}
        c_\mu
        R_{\mu \boldsymbol{\ell}}
        ( \boldsymbol{a}_{\boldsymbol{\ell} }, \boldsymbol{k}_{\mathrm{c} \boldsymbol{\ell}} )
\end{align}
that depends on the target MO via the coefficients $\{ c_\mu \}_\mu.$

\section{Derivation of Eq.~(\ref{ampl_encoding_of_GMO:width_deriv_of_ovl_for_optimal_coeffs})}
\label{sec:width_deriv_of_ovl_for_optimal_coeffs}

For any one $\widetilde{\boldsymbol{d}}$ of the eigenvectors for
Eq.~(\ref{ampl_encoding_of_GMO:eig_prob_for_lc_coeffs}),
we differentiate the both sides of the normalization condition in
Eq.~(\ref{ampl_encoding_of_GMO:normalization_cond_of_trial})
with respect to $a_{\nu \ell }$ to obtain
\begin{gather}
        \frac{\partial \widetilde{\boldsymbol{d}}}{\partial a_{\nu \ell } }
        \cdot
        S ( \boldsymbol{a}, \boldsymbol{k}_{\mathrm{c}} )
        \widetilde{\boldsymbol{d}}
        +
        \widetilde{\boldsymbol{d}}
        \cdot
        \frac{\partial S ( \boldsymbol{a}, \boldsymbol{k}_{\mathrm{c}} ) }{\partial a_{\nu \ell } }
        \widetilde{\boldsymbol{d}}
        +
        \widetilde{\boldsymbol{d}}
        \cdot
        S ( \boldsymbol{a}, \boldsymbol{k}_{\mathrm{c}} )
        \frac{\partial \widetilde{\boldsymbol{d}}}{\partial a_{\nu \ell } }
    =
        0
        .
    \label{ampl_encoding_of_GMO:deriv_normalization_cond_of_trial}
\end{gather}
The derivative of the fidelity can be obtained by
differentiating
Eq.~(\ref{ampl_encoding_of_GMO:ovl_btwn_MO_and_trial_using_eigenvector})
with respect to $a_{\nu \ell }$ as
\begin{align}
    \frac{
        \partial
        F ( \widetilde{\boldsymbol{d}}, \boldsymbol{a}, \boldsymbol{k}_{\mathrm{c}} )
    }{\partial a_{\nu \ell } }
    &=
        \frac{\partial \widetilde{\boldsymbol{d}} }{\partial a_{\nu \ell } }
        \cdot
        G ( \boldsymbol{a}, \boldsymbol{k}_{\mathrm{c}} )
        \widetilde{\boldsymbol{d}}
        +
        \widetilde{\boldsymbol{d}}
        \cdot
        \frac{
            \partial
            G ( \boldsymbol{a}, \boldsymbol{k}_{\mathrm{c}} )
        }{\partial a_{\nu \ell } }
        \widetilde{\boldsymbol{d}}
        +
        \widetilde{\boldsymbol{d}}
        \cdot
        G ( \boldsymbol{a}, \boldsymbol{k}_{\mathrm{c}} )
        \frac{\partial \widetilde{\boldsymbol{d}} }{\partial a_{\nu \ell } }
        -
        \frac{\partial P (\boldsymbol{a}, \boldsymbol{k}_{\mathrm{c}})}{\partial a_{\nu \ell} }
    \nonumber \\
    &=
        \frac{\partial \widetilde{\boldsymbol{d}} }{\partial a_{\nu \ell } }
        \cdot
        \kappa
        S ( \boldsymbol{a}, \boldsymbol{k}_{\mathrm{c}} )
        \widetilde{\boldsymbol{d}}
        +
        \widetilde{\boldsymbol{d}}
        \cdot
        \frac{
            \partial
            G ( \boldsymbol{a}, \boldsymbol{k}_{\mathrm{c}} )
        }{\partial a_{\nu \ell } }
        \widetilde{\boldsymbol{d}}
        +
        \kappa
        \widetilde{\boldsymbol{d}}
        \cdot
        S
        ( \boldsymbol{a}, \boldsymbol{k}_{\mathrm{c}} )
        \frac{\partial \widetilde{\boldsymbol{d}} }{\partial a_{\nu \ell } }
        -
        \frac{\partial P (\boldsymbol{a}, \boldsymbol{k}_{\mathrm{c}})}{\partial a_{\nu \ell} }
    \nonumber \\
    &=
        \widetilde{\boldsymbol{d}}
        \cdot
        \left(
            \frac{
                \partial G ( \boldsymbol{a}, \boldsymbol{k}_{\mathrm{c}} )
            }{\partial a_{\nu \ell } }
            -
            \kappa
            \frac{
                \partial S ( \boldsymbol{a}, \boldsymbol{k}_{\mathrm{c}} )
            }{\partial a_{\nu \ell } }
        \right)
        \widetilde{\boldsymbol{d}}
        -
        \frac{\partial P (\boldsymbol{a}, \boldsymbol{k}_{\mathrm{c}})}{\partial a_{\nu \ell} }
        ,
    \label{ampl_encoding_of_GMO:deriv_of_ovl_wrt_width}
\end{align}
where we used
Eqs.~(\ref{ampl_encoding_of_GMO:eig_prob_for_lc_coeffs})
and (\ref{ampl_encoding_of_GMO:deriv_normalization_cond_of_trial})
for obtaining the second and third equalities, respectively.
As a special case,
we substitute the optimal core tensor
$\boldsymbol{d} ( \boldsymbol{a}, \boldsymbol{k}_{\mathrm{c}} )$
given by
Eq.~(\ref{ampl_encoding_of_GMO:optimal_d_for_fixed_widths_and_centers})
into Eq.~(\ref{ampl_encoding_of_GMO:deriv_of_ovl_wrt_width})
to obtain
Eq.~(\ref{ampl_encoding_of_GMO:width_deriv_of_ovl_for_optimal_coeffs}).

\section{CNOT gate counts based on UCRs}
\label{sec:num_cnot_based_on_ucrs}

\subsection{Generic amplitude encoding}

A UCR for a generic $n$-qubit system is defined as $2^{n - 1}$ single-qubit rotations on one qubit controlled by the other $n - 1$ qubits
where every possible control pattern (being controlled or anticontrolled) appears only once.
This UCR is known to be implemented by using $2^{n - 1}$ CNOT gates \cite{bib:5693}.
Given this fact, the CNOT gate count in the generic amplitude encoding circuit in Fig.~\ref{fig:circuit_lcu_ampl} is calculated as
\begin{align}
    N_{\mathrm{C} X} (U_{\mathrm{amp}} [\boldsymbol{c}])
    =
        \sum_{k = 1}^{n - 1}
            2^k
    =
        2^n - 2
        .
    \label{ampl_encoding_of_GMO:num_CNOT_for_LCU_using_unif_ctrl_rot}
\end{align}

\subsection{Tucker-form state}

The original implementation of the part of 
$U_{\mathrm{S-ph}} [\boldsymbol{a}, \boldsymbol{k}_{\mathrm{c}}]$
used in Fig.~\ref{fig:circuit_tucker}
for generating $n_{\mathrm{L} \nu}$ 1D LFs at the origin for the $\nu$ direction $(\nu = x, y, z)$ is shown in Fig.~\ref{fig:circuit_unif_ctrl_rot_for_Slater}(a).
This partial circuit consists of multiply controlled multiple $y$ rotations.
By defining 
$U_{\mathrm{S} \nu}^{(k)} \ (k = 0, \dots, n_{qe} - 1)$
for a single qubit and the $n_{\mathrm{A} \nu}$ Lorentzian ancillae
as shown in Fig.~\ref{fig:circuit_unif_ctrl_rot_for_Slater}(b)
and rearranging the controlled rotations in
Fig.~\ref{fig:circuit_unif_ctrl_rot_for_Slater}(a) appropriately,
the circuit is expressed using the product of operations $U_{\mathrm{S} \nu}^{(k)}.$
Also, the partial circuit for giving rise to the phase factors of SFs can undergo similar rearrangement of the phase gates.
The $\nu$ direction part of
$U_{\mathrm{S-ph}} [\boldsymbol{a}, \boldsymbol{k}_{\mathrm{c}}]$
can thus be implemented by the circuit shown in 
Fig.~\ref{fig:circuit_unif_ctrl_rot_for_Slater}(c). 
Since each $U_{\mathrm{S} \nu}^{(k)}$ forms a UCR, we have
$
N_{\mathrm{C} X} (U_{\mathrm{S} \nu}^{(k)})
=
2^{ n_{\mathrm{A} \nu} }
.
$
Since each $U_{\mathrm{shift} \nu}^{(k)}$ is multiply controlled phase gates for an $(n_{\mathrm{A} \nu} + 1)$-qubit system,
it is a diagonal unitary.
This can be implemented by using the technique in
Ref.~\cite{bib:5352} with
$
N_{\mathrm{C} X} (U_{\mathrm{Shift} \nu}^{(k)})
=
2^{ n_{\mathrm{A} \nu} + 1} - 2
.
$
When we adopt the implementation of consecutive CNOTs in Ref.~\cite{bib:6809}, we have
$
N_{\mathrm{C} X} (\mathrm{CNOT}^{\otimes n_{q e}})
=
2 n_{q e} - 3
.
$
The CNOT gate count for generating the SFs and their phase factors is thus calculated to be
\begin{align}
    N_{\mathrm{C} X}
    ( U_{\mathrm{S-ph}} [\boldsymbol{a}, \boldsymbol{k}_{\mathrm{c}}] )
    &=
        -9
        +
        3 n_{q e}
        \sum_{\nu=x,y,z}
            2^{ n_{\mathrm{A} \nu} }
        .
    \label{ampl_encoding_of_GMO:num_cnots_in_SFs_and_phases}
\end{align}

The CNOT gate count for the amplitude encoding of the core tensor is,
similarly to
Eq.~(\ref{ampl_encoding_of_GMO:num_CNOT_for_LCU_using_unif_ctrl_rot}),
calculated to be
\begin{align}
    N_{\mathrm{C} X} (U_{\mathrm{amp}} [ \boldsymbol{d} ])
    =
        2^{n_{\mathrm{A}}^{(\mathrm{L})}} - 2
    .
\end{align}
From this and Eq.~(\ref{ampl_encoding_of_GMO:num_cnots_in_SFs_and_phases}),
the gate count in the state preparation circuit for a Tucker-form state without QFT is
\begin{align}
    N_{\mathrm{C} X}
    (\mathcal{C}_{\mathrm{Tucker}}^{(\mathrm{prob})})
    &=
        2^{n_{\mathrm{A}}^{(\mathrm{L})}} 
        -11
        +
        3 n_{q e}
        \sum_{\nu}
            2^{n_{\mathrm{A} \nu} }
        .
\end{align}

\begin{figure*}
\begin{center}
\includegraphics[width=16cm]{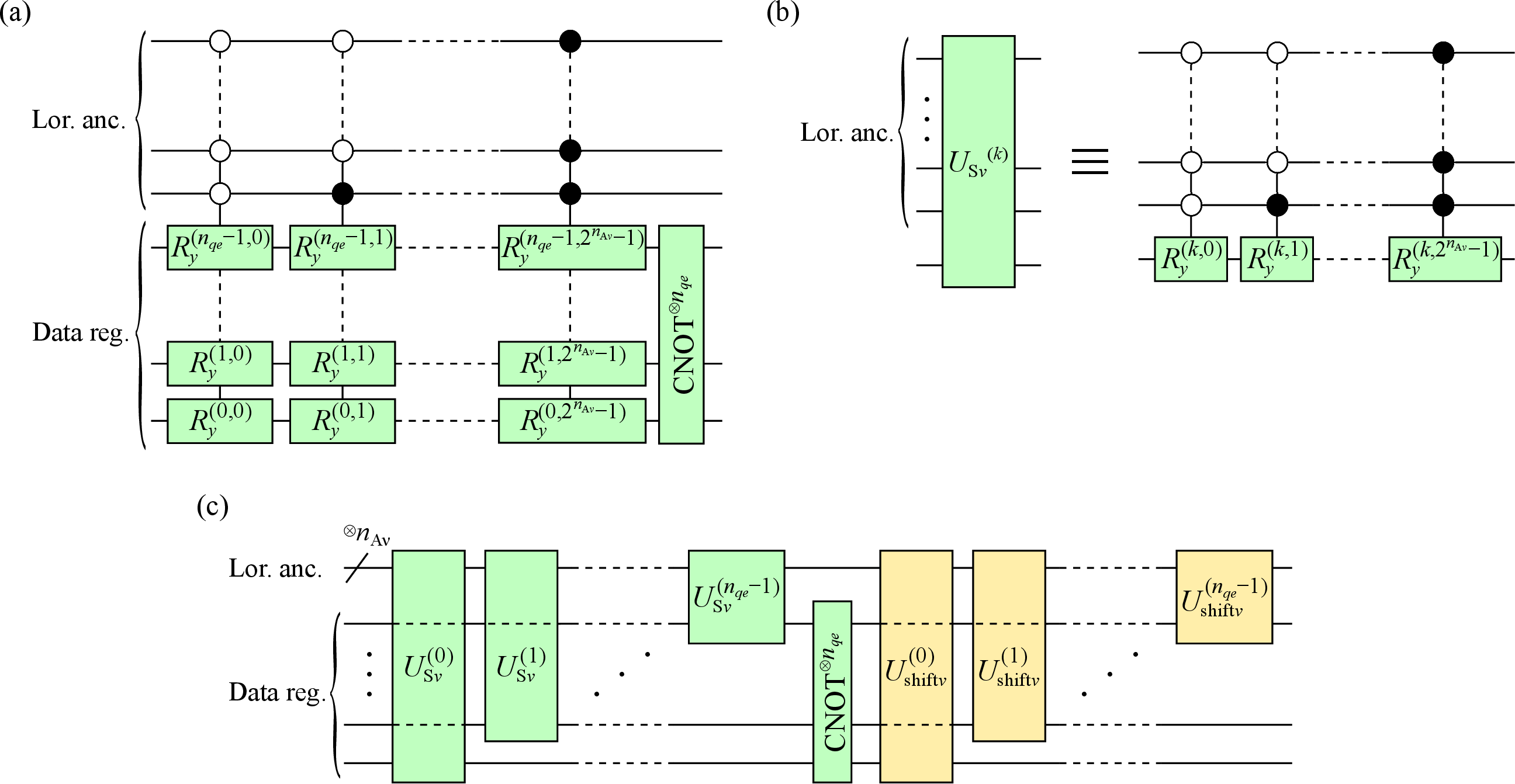}
\end{center}
\caption{
(a)
The original implementation of the part of 
$U_{\mathrm{S-ph}} [\boldsymbol{a}, \boldsymbol{k}_{\mathrm{c}}]$
for generating $n_{\mathrm{L} \nu}$ 1D LFs at the origin for the $\nu$ direction $(\nu = x, y, z)$.
$R_y^{(k,\ell)}$ in the figure is the $y$ rotation gate acting on the $k$th qubit in the data register to generate the $\ell$th SF.
(b)
Definition of 
$U_{\mathrm{S} \nu}^{(k)} \ (k = 0, \dots, n_{qe} - 1)$
for a single qubit and the $n_{\mathrm{A} \nu}$ Lorentzian ancillae.
By rearranging the controlled rotations in (a) appropriately,
the circuit is expressed using the product of $U_{\mathrm{S} \nu}^{(k)}$ operations.
Also, the partial circuit for giving rise to the phase factors of SFs can undergo similar rearrangement of the phase gates, leading to the circuit depicted in (c), that still implements the $\nu$ direction part of $U_{\mathrm{S-ph}} [\boldsymbol{a}, \boldsymbol{k}_{\mathrm{c}}].$
}
\label{fig:circuit_unif_ctrl_rot_for_Slater}
\end{figure*}

\subsection{Canonical-form state}

The CNOT gate count for the amplitude encoding of the normalized canonical coefficients is,
from Eq.~(\ref{ampl_encoding_of_GMO:num_CNOT_for_LCU_using_unif_ctrl_rot}),
$
N_{\mathrm{C} X} (U_{\mathrm{amp}} [\widetilde{\boldsymbol{\lambda}}])
=
2^{n_{\mathrm{A}}^{(\mathrm{c})}} - 2
.
$
The part of $U_{\mathrm{canon, amp}}$ for amplitude encoding of the normalized canonical tensors can be expressed by using UCRs as well as 
the case of
$U_{\mathrm{S-ph}} [\boldsymbol{a}, \boldsymbol{k}_{\mathrm{c}}].$
Specifically, Fig.~\ref{fig:circuit_unif_ctrl_rot_for_canon}(a) shows
the original implementation of the part of $U_{\mathrm{canon,amp}}$
in Fig.~\ref{fig:circuit_canonical}
for amplitude encoding of the normalized canonical tensors $\boldsymbol{u}^{(\nu)}$ for the $\nu$ direction.
This circuit acts on the $n_{\mathrm{A}}^{(\mathrm{c})}$ canonical ancillae and the $n_{\mathrm{A} \nu}$ Lorentzian ancillae.
We rearrange the controlled operations appropriately to obtain the equivalent circuit shown in
Fig.~\ref{fig:circuit_unif_ctrl_rot_for_canon}(b),
that is the product of operations
$U_{\mathrm{canon} \nu}^{(k)} [\boldsymbol{u}^{(\nu)}].$
Since each
$U_{\mathrm{canon} \nu}^{(k)} [\boldsymbol{u}^{(\nu)}]$
is a UCR, it can be implemented with
$
N_{\mathrm{C} X} (U_{\mathrm{canon} \nu}^{(k)} [\boldsymbol{u}^{(\nu)}])
=
2^{ n_{\mathrm{A}}^{(\mathrm{c})} + k }
.
$
The CNOT gate count of the circuit in
Fig.~\ref{fig:circuit_unif_ctrl_rot_for_canon}(b)
is thus given by
\begin{align}
    \sum_{k = 0}^{n_{\mathrm{A} \nu} - 1}
        2^{ n_{\mathrm{A}}^{(\mathrm{c})} + k }
    =
        2^{ n_{\mathrm{A}}^{(\mathrm{c})} }
        (2^{n_{\mathrm{A} \nu} } - 1)
    .
\end{align}
This fact enables us to implement the partial circuit with
\begin{align}
    N_{\mathrm{C} X} (U_{\mathrm{canon, amp}})
    &=
        2^{n_{\mathrm{A}}^{(\mathrm{c})}} - 2
        +
        \sum_{\nu = x, y, z}
        2^{ n_{\mathrm{A}}^{(\mathrm{c})} }
        (2^{n_{\mathrm{A} \nu} } - 1)
        .
\end{align}
From this and Eq.~(\ref{ampl_encoding_of_GMO:num_cnots_in_SFs_and_phases}),
the gate count in the state preparation circuit for a canonical-form state without QFT is
\begin{align}
    N_{\mathrm{C} X}
    (\mathcal{C}_{\mathrm{canon}}^{(\mathrm{prob})} )
    &=
        -
        2^{ n_{\mathrm{A}}^{(\mathrm{c})} + 1}
        -
        11
        +
        (3 n_{qe} + 2^{ n_{\mathrm{A}}^{(\mathrm{c})} } )
        \sum_\nu
        2^{n_{\mathrm{A} \nu}}
    .
\end{align}

\begin{figure*}
\begin{center}
\includegraphics[width=14cm]{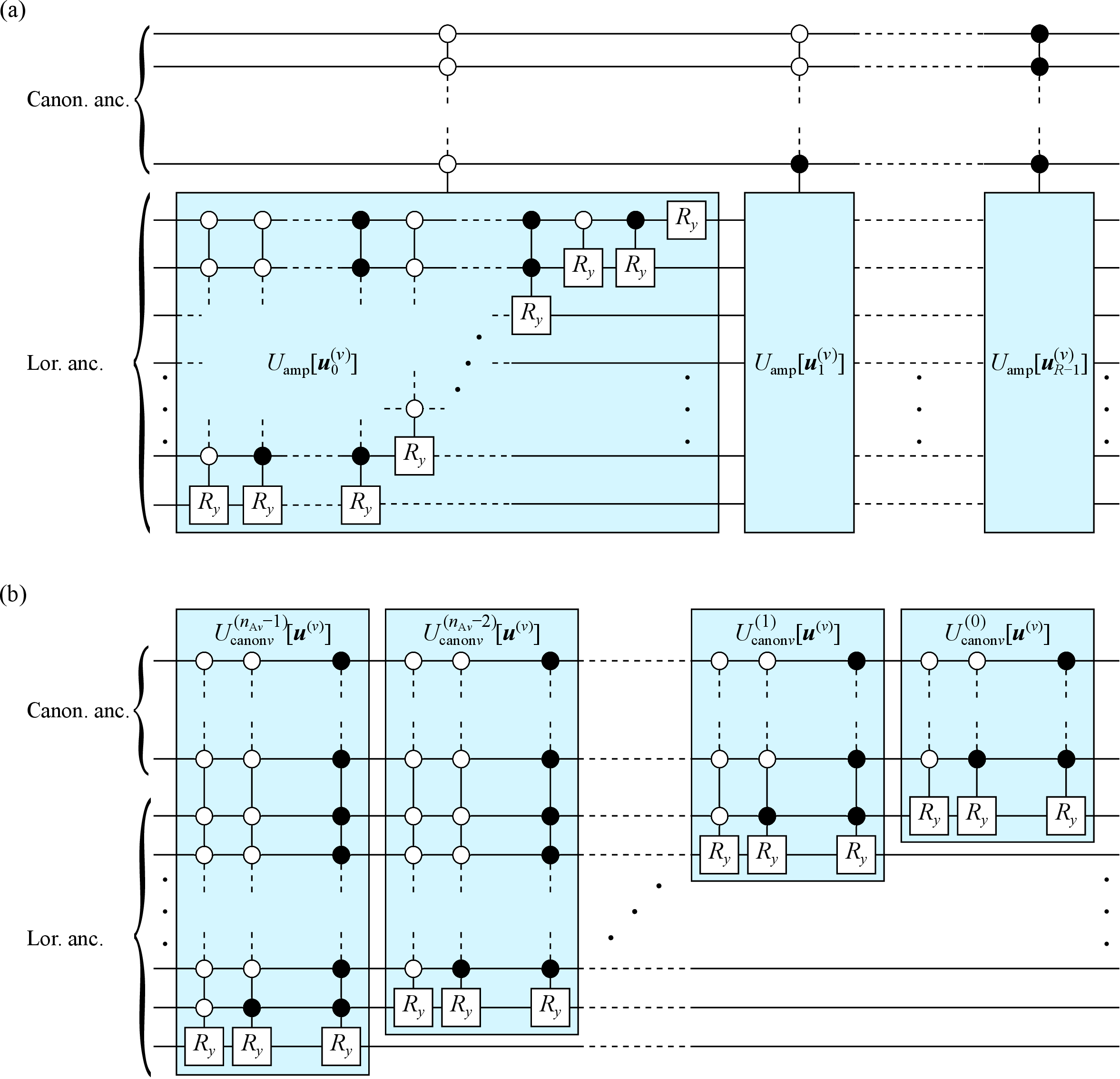}
\end{center}
\caption{
(a)
The original implementation of the part of $U_{\mathrm{canon,amp}}$
for amplitude encoding of the normalized canonical tensors $\boldsymbol{u}^{(\nu)}$ for the $\nu$ direction ($\nu = x, y, z$).
By rearranging the controlled operations appropriately,
the equivalent circuit in (b) is obtained.
Each 
$U_{\mathrm{canon} \nu}^{(k)} [\boldsymbol{u}^{(\nu)}] \ (k = 0, \dots, n_{\mathrm{A} \nu} - 1)$
is a UCR having $n_{\mathrm{A}}^{(\mathrm{c})} + k$ control bits.
}
\label{fig:circuit_unif_ctrl_rot_for_canon}
\end{figure*}

\section{Simple explanation of a low success probability for an antibonding overlap}
\label{sec:success_prob_lowered_by_antibonding}

In order for discussion to be simple,
we consider here two atoms $A$ and $B$ in 1D space.
We assume that we have a trial state
$
| \psi \rangle
=
(\cos \theta U_A + \sin \theta U_B) | 0 \rangle
$
as a linear combination of the LFs with an angle parameter $\theta.$
$U_A$ and $U_B$ are the unitaries that generate the discrete LFs centered at the position of $A$ and $B,$ respectively.
The success probability for generating this state via LCU for the two unitaries is
$\mathbb{P} = 1/2 + \sin (2 \theta) \langle L_A | L_B \rangle/2.$
Since the LFs are positive from their definitions, we can write
$\langle L_A | L_B \rangle = 1 - \Delta$ with some $\Delta \geq 0.$

If the trial state is of bonding nature, we can parametrize it as $\theta = \pi/4 + \delta$ with a small $\delta.$
The success probability in this case is
$\mathbb{P}_{\mathrm{bonding}} \approx 1 - \Delta/2 - \delta^2 ( 1 - \Delta).$
If the trial state is of antibonding nature, on the other hand,
the success probability with an angle parameter $\theta = -\pi/4 + \delta$ is
$\mathbb{P}_{\mathrm{antibonding}} \approx \Delta/2 + \delta^2 ( 1 - \Delta).$
This analysis tells us that the success probability for a bonding MO between $A$ and $B$ tends to be higher than that for an antibonding MO between the same atoms.
In particular,
as the two centers get closer to each other, $\Delta$ becomes smaller and the tendency is more pronounced.
This discussion may apply basically to generic trial states consisting of more than two LFs in 3D space.

\section{Canonical-form states for an H$_2$O molecule}
\label{sec:canon_states_of_H2O}

For each of MOs $2 a_1, 1 b_2, 3 a_1,$ and $1 b_1$ of an H$_2$O molecule,
we provide here the plots of the states $| \phi_{\mathrm{canon}, r}^{(\nu)} \rangle \ (\nu = x, y, z)$ that constitute the rank-$R$ canonical-form state $| \phi_{\mathrm{canon}} \rangle,$
as shown in Fig.~\ref{fig:h2o_canon}(a)-(d).
The success probabilities of state preparation for the canonical-form states are plotted in Fig.~\ref{fig:h2o_canon}(e).

\begin{figure*}
\begin{center}
\includegraphics[width=15cm]{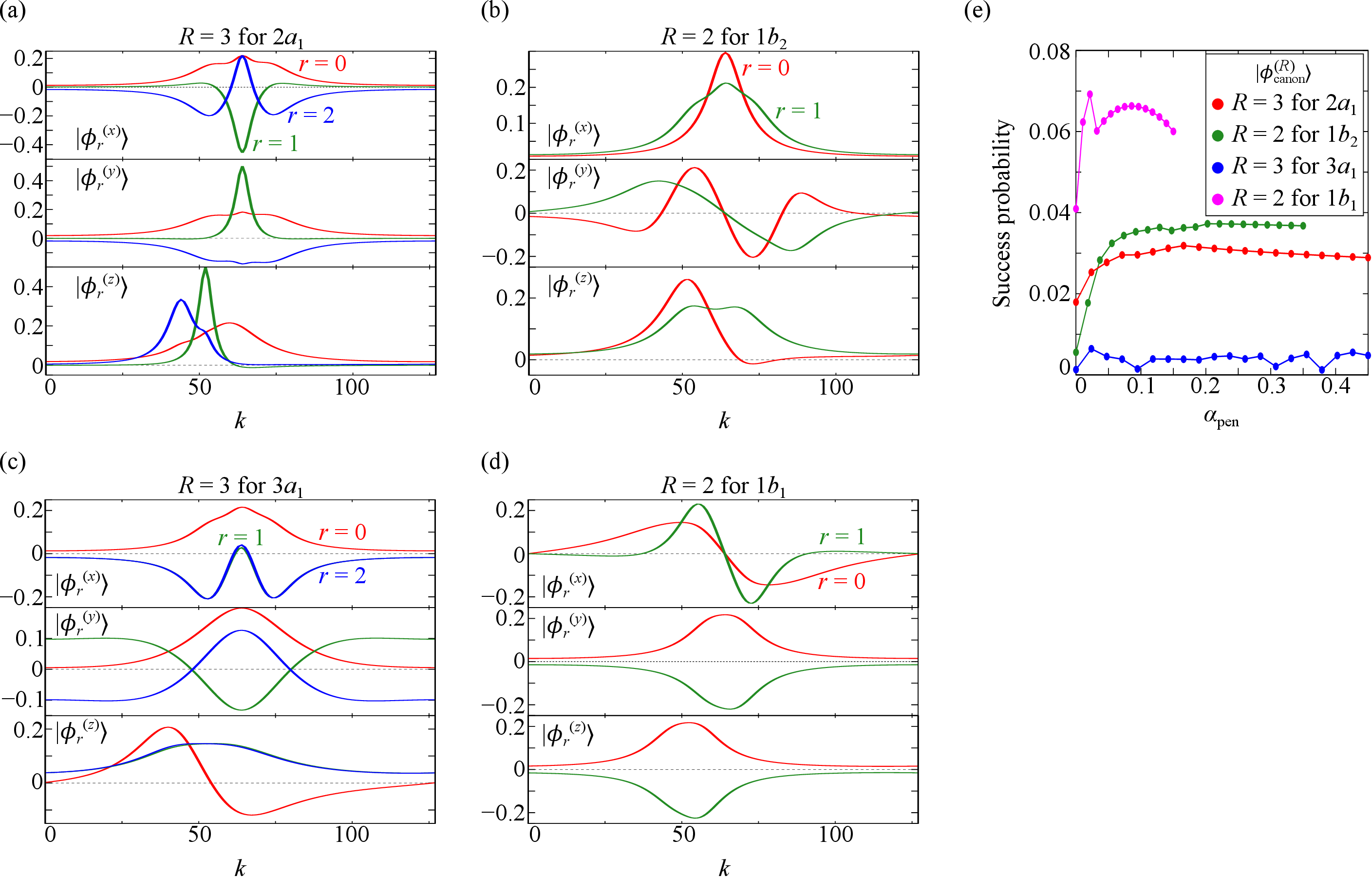}
\end{center}
\caption{
(a)
$| \phi_r^{(\nu)} \rangle$ states projected along the computational basis $| k \rangle_{n_{q e}}$ for directions $\nu$.
They constitute the rank-3 canonical-form state for the $2 a_1$ MO.
Similar plots for the $1 b_2, 3 a_1,$ and $1 b_1$ MOs are shown in (b), (c), and (d), respectively.
(e)
Success probabilities of state preparation for the canonical-form states.
}
\label{fig:h2o_canon}
\end{figure*}

\end{widetext}

\bibliography{ref}

\end{document}